\documentclass[prb,twocolumn,flotfix]{revtex4-2}

\usepackage{float}
\usepackage{placeins}
\usepackage{latexsym}
\usepackage{graphicx}

\usepackage{subfigure}    

\usepackage{amsmath}
\usepackage{color}
\usepackage{xcolor}
\usepackage{soul}

\usepackage{relsize}
\usepackage{placeins}
\usepackage{amssymb}
\usepackage{amsthm}
\usepackage{amsfonts}
\usepackage{braket}
\usepackage{latexsym}
\usepackage{graphicx}
\usepackage{grffile}
\usepackage{hyperref}
\usepackage{multirow}

\begin{document}
	
\

\title{Towards understanding the electronic structure of the simpler members of two-dimensional halide-perovskites} 

\author{Efstratios Manousakis} 
\affiliation{Department  of  Physics, Florida  State  University,  Tallahassee,  FL  32306-4350,  USA\\
  and Department of Physics, National and Kapodistrian University of Athens,
Penepistimioupolis Zografos, 157 84 Athens, Greece}
\date{\today} 

\begin{abstract}
  In this paper we analyze the band-structure of 
  two-dimensional (2D) halide perovskites by considering structures related to the simpler case of the series,
  (BA)$_2$PbI$_4$, in which PbI$_4$ layers are intercalated
  with butylammonium (BA=CH$_3$(CH$_2$)$_3$NH$_3$) organic ligands.
   We use density-functional-theory (DFT) based
   calculations and tight-binding (TB) models aiming to discover a simple description of the bands within 1 eV below the valence-band maximum
    and 2 eV above the conduction-band minimum, which, including the energy gap, is about a $\Delta E =5$ eV energy range.
   The bands in this  $\Delta E$ range are those expected to contribute to the
   transport phenomena, photoconductivity and light-emission in the visible spectrum, at room and low temperature.
  We find that the atomic orbitals of the
  butylammonium chains have negligible contribution to the Bloch states which form the conduction and valence bands in the above defined $\Delta E$ range.
   Our calculations reveal a rather universal, i.e., independent
   of the intercalating BA, rigid-band picture
  inside the above $\Delta E $ range characteristic of the layered perovskite
  ``matrix'' (i.e., PbI$_4$ in our example).
  Besides demonstrating the above conclusion,  the main goal of
  this paper is to find accurate TB
  models which capture the essential features of the DFT bands
  in this $\Delta E$ range.
  First, we ignore electron hopping along the $c$-axis and
  the octahedral distortions  and this increased symmetry (from C$_2$ to C$_4$)
  halves the Bravais-lattice unit-cell size and the 
  Brillouin zone  unfolds to a 45$^{\circ}$ rotated square and this
  allows some analytical handling of the 2D TB-Hamiltonian.
  The Pb $6s$ and I $5s$ orbitals are far away from the
  above $\Delta E$ range and, thus, we integrate them out to obtain an effective
  model which only includes hybridized Pb $6p$ and I $5p$ states.
 Our TB-based treatment a) provides a good quantitative description of the
  DFT band-structure, b) helps us conceptualize the complex electronic structure
  in the family  of these materials in a simple way and c) yields
  the one-body part to be combined with appropriately screened electron
  interaction to describe many-body effects,  such as,
  excitonic bound-states.
 
\end{abstract}
\maketitle

\section{Introduction} 

The discovery and production of semiconducting superlattices has led to a significant advancement in solid-state physics and electronic technology.
Further development, however, has been hindered by the infrastructure-demanding time-consuming fabrication procedure for precisely assembling the nanometer-scale structures
in such artificial material structures.
Rather recently, a new class of superlattices has been discovered, known as Ruddlesden-Popper (RP) two-dimensional (2D) halide-perovskites, that can be self-assembled using wet-chemistry
synthesis\cite{Stoumpos2016,Mitzi1996}.
The interest in this type of materials is not just because of the simplicity of the synthesis process, but more importantly, the interest is due to the fact that they are clean and maintain the periodic structure
at the atomic level.

\begin{figure}[!htb]
\includegraphics[width=1.0\linewidth]{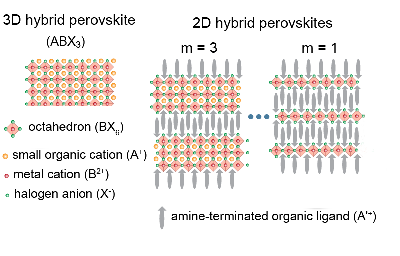} 
\caption{Illustration of the relationship of the Ruddlesden-Popper perovskite structures
to the three-dimensional (3D) perovskite ABX$_3$. See text for details.}
\label{2d-perovskite}
\end{figure}
These 2D halide-perovskites are analogues to the oxide perovskites
described by Ruddlesden and Popper\cite{RPopper}, from where they gain their name, and they form self-assembled superlattice structures\cite{Stoumpos2016,Gao2019,doi:10.1126/science.aac7660,Blancon2020}.
Fig.~\ref{2d-perovskite} illustrates the relationship of the Ruddlesden-Popper perovskite structures
to the three-dimensional (3D) perovskite ABX$_3$, where A B and X  are,  a small
organic cation (such as methylammonium), a metal cation and a halogen anion respectively.
The bulk perovskite ABX$_3$ is figuratively sliced apart to form a multilayer
sandwich in which 
amine-terminated organic ligands A$^{\prime}$ (such as, butylammonium) are ``inserted'' in between
every m (m=1,2,...) adjacent halide-perovskite layers (see Fig.~\ref{2d-perovskite}) in the formula A$^{\prime}_2$A$_{\mathrm{m}-1}$B$_m$X$_{3 \mathrm{m}+1}$. An example of this series is the family\cite{Stoumpos2016,Mitzi1996}
(BA)$_2$(MA)$_{\mathrm{m}-1}$Pb$_{\mathrm{m}}$I$_{3\mathrm{m}+1}$ with BA=CH$_3$(CH$_2$)$_3$NH$_3$ and MA=CH$_3$NH$_3$ (m = 1, 2, 3, 4, ...). These Ruddlesden-Popper
hybrid perovskites come with inherent
low dimensionality and highly ordered periodic nanostructures. The amine-terminated organic
ligands essentially cleave the 3D perovskite lattices into two-dimensional sheets,
forming alternating layers of organic and inorganic supercells along the $c$-axis crystal orientation.
The formation of such layered structures is thermodynamically
favorable, which makes it possible to obtain Ruddlesden-Propper hybrid perovskites conveniently
using wet-chemistry synthesis.

The family of 3D organic halide-perovskites has received
tremendous attention during the past decade due to their promising
optoelectronic and solar cell applications.
Decreasing the
dimensionality from 3D to 2D, more versatile organic cations can be incorporated
as templates to produce new structures\cite{Manser2016}.
The properties of these reduced dimensionality
semiconductors are less widely studied but results of many such studies have started to appear in the literature rapidly. 
Among the many fascinating properties of these materials is white-light emission\cite{Thirumal2017}, where the broad emission comes from the transient photoexcited states generated by self-trapped excitons. 
Furthermore, these materials have a significant flexibility in tuning their optoelectronic properties by varying the number of perovskite layers and by choosing an
appropriate organic ligand. This makes them particularly suitable for
several  photovoltaic applications\cite{D1CS00106J} and as
light emitters.\cite{Thirumal2017,https://doi.org/10.1002/advs.202102689,Ban2018,Zhang2021,doi:10.1063/5.0047616}
Fast pump-probe spectroscopies have also revealed useful information
about the carrier dynamics and recombination of these RP perovskite materials\cite{en14030708,Milot2016,Cho2020,https://doi.org/10.1002/adom.201900971},
as well as effects of excitonic many-body interactions\cite{Wu2015}.

Examining these materials from a different view angle, these RP perovskites can potentially provide a new playground for fundamental physics. Namely, these nearly ideal 2D structures add another interesting family
of materials to a growing list of interesting 2D materials, such as, Graphene and its variants and its cousin materials, transition metal dichalcogenides, etc,  which can host a variety of interesting phenomena, and
potentially new phases of matter. 
Understanding the physics beyond single-electron phenomena, and in order
to make further progress, requires understanding the band structure of these
materials at a deeper yet simpler level than the complex multi-band picture provided by the DFT calculation.
However, the number of atoms in a single unit cell of the Bravais lattice
is very large. For example, the unit-cell of the  Bravais lattice of the m=1
structure, which is the subject of the present paper, contains 156 atoms.
Therefore, the complexity of the atomic structure of the
material, which is reflected in its band-structure, might be a reason for not finding them appealing for theoretical investigations of potential novel
phenomena. Namely, at first sight, it might seem a hopeless task to try to find a simple picture to describe the electronic structure.

There are numerous publications where DFT and related techniques have been
applied in order to understand various aspects of these and related materials.
The aim of the present paper is not to add another such study, but
rather to analyze the known complex band-structure\cite{PhysRevB.67.155405} of (BA)$_2$PbI$_4$ and related materials
and find a simple (and, if possible, analytical) and accurate way to describe the origin of its features. Interestingly, we find that, in the simplest case
of (BA)$_2$PbI$_4$, it is possible
to achieve this goal. First, we show that a rigid-band description is
accurate for such materials. For example, we find that
all the bands within 2 eV below the valence-band maximum (VBM) and 2 eV above the conduction-band minimum (CBM) have
negligible contribution from the atomic orbitals contained in the amine-terminated organic ligands. This is not to say that these larger organic ligands do not play
a significant role in various aspects of the crystal formation.
For example, the choice of these
organic spacers is important in achieving good quality crystallization\cite{WU2021582,CAO2022101394}
and allows optimization of the film quality\cite{D1CS00106J}, and in achieving the 
crystal orientation
and the stability of the system\cite{Chen2018,Gao2019}
However, once the structure is formed and the positions of all the metal and halogen atoms
are given, our calculation illustrates that the Bloch states with energy
within 1 eV below the VBM and 2 eV above the CBM,
have negligible projection to the atomic orbitals of these large organic ligands.
The role of these larger butylammonium organic ligands is simply to act as a charge reservoir
which fill completely the highest occupied bands
making the material an insulator. 

The second part of the present paper is to uncover a simple picture of the band-structure responsible for most of the optoelectronic response.
We consider a 2D TB model, which ignores
the small octahedra distortions and this allows us to reduce the size of
the unit-cell by a factor of two, a fact that doubles the Brillouin-zone (BZ)
by unfolding it because of symmetry. This reduces the TB Hamiltonian
to a $12 \times 12$ matrix for each point in the BZ. The most important
conduction and valence bands are obtained as  a hybridization
of mostly metal-ion $p$ and halogen $p$ orbitals. In the case of our example, (BA)$_2$PbI$_4$,
hybridization occurs between Pb and I $p$ orbitals. To obtain the correct dispersion
of the highest valence-band, we also need to involve the role of the $sp$ hybridization
between the metal-ion $s$ orbital and the $p$ orbital of the halogen atoms that
sit at the octahedra corners. Finally, we offer a simple analytic description
of the band-structure by integrating out this metal-ion $s$ orbital as it is 
energetically well below the Fermi energy.

The paper is organized as follows. In Sec.~\ref{c-and-b-structure} we present the results of our DFT calculation (including the projection of the Bloch states to atomic orbitals)  for those crystalline structures which we think are relevant for the point to be made.
In Sec.~\ref{sec:TB} we detail our tight-binding model and how it fits with the DFT results of the bands and the orbital character of the
Bloch wavefunctions. In Sec.~\ref{analytical} we present the analytical model that gives a good approximation to the bands near the Fermi level
and we also give the final fit or our TB model to the DFT bands.
In Sec.~\ref{discussion} we present our final remarks and conclusions of our study.

\section{The crystal and band structure}
\label{c-and-b-structure}
\subsection{Crystal structure}
\begin{figure*}[!htb]
  \includegraphics[width=0.28\linewidth]{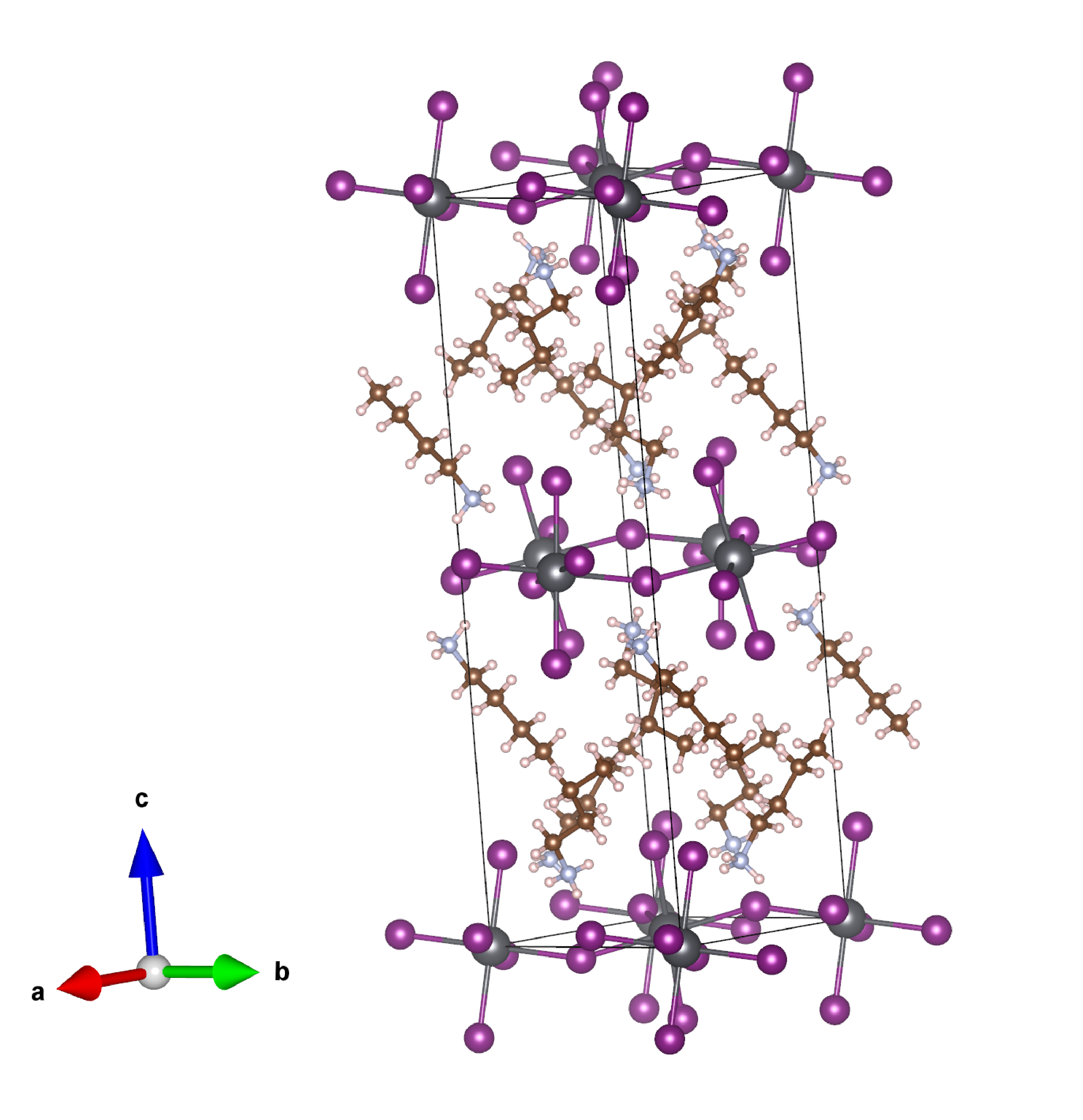}
  \includegraphics[width=0.28\linewidth]{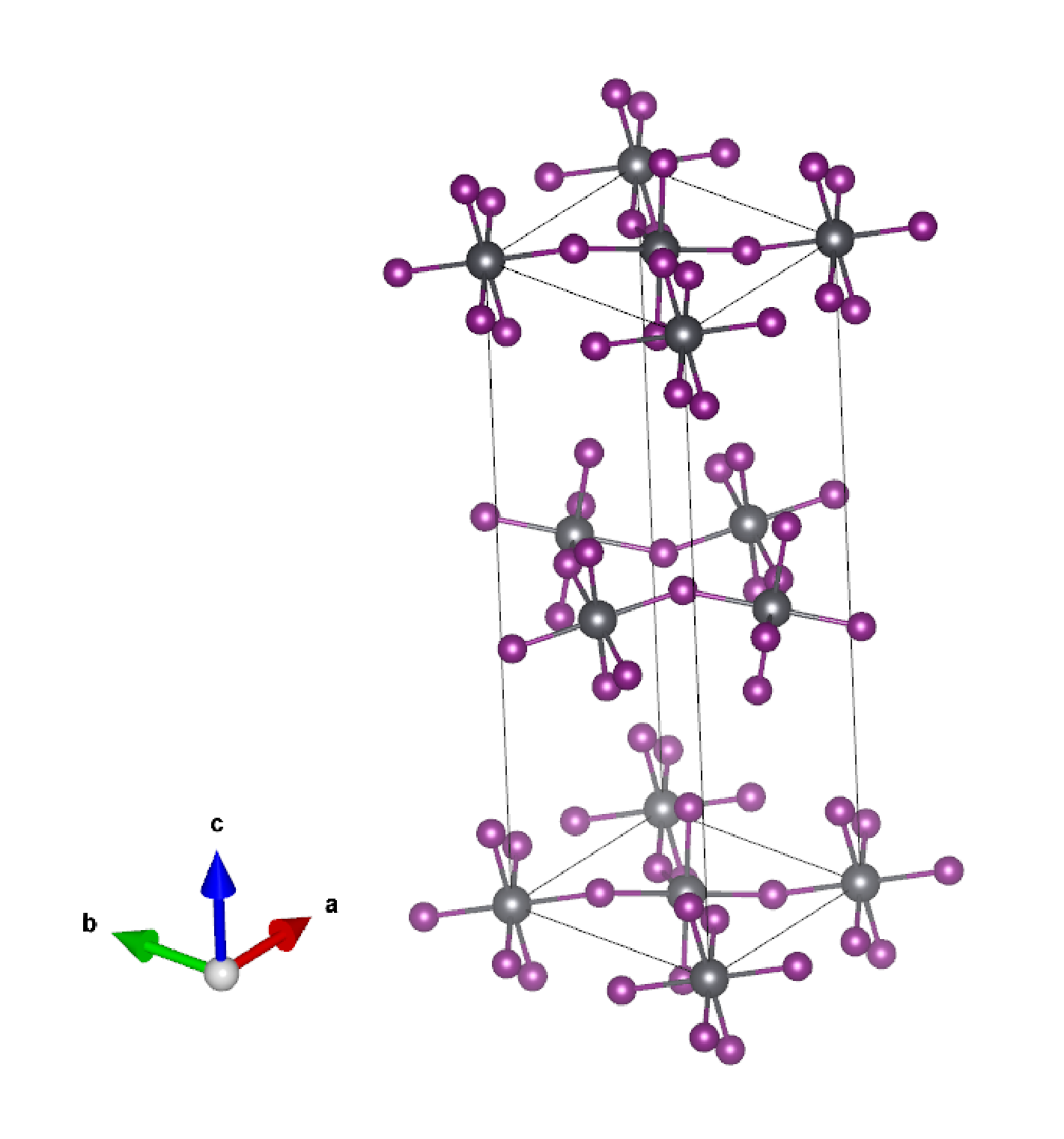} 
  \includegraphics[width=0.4\linewidth]{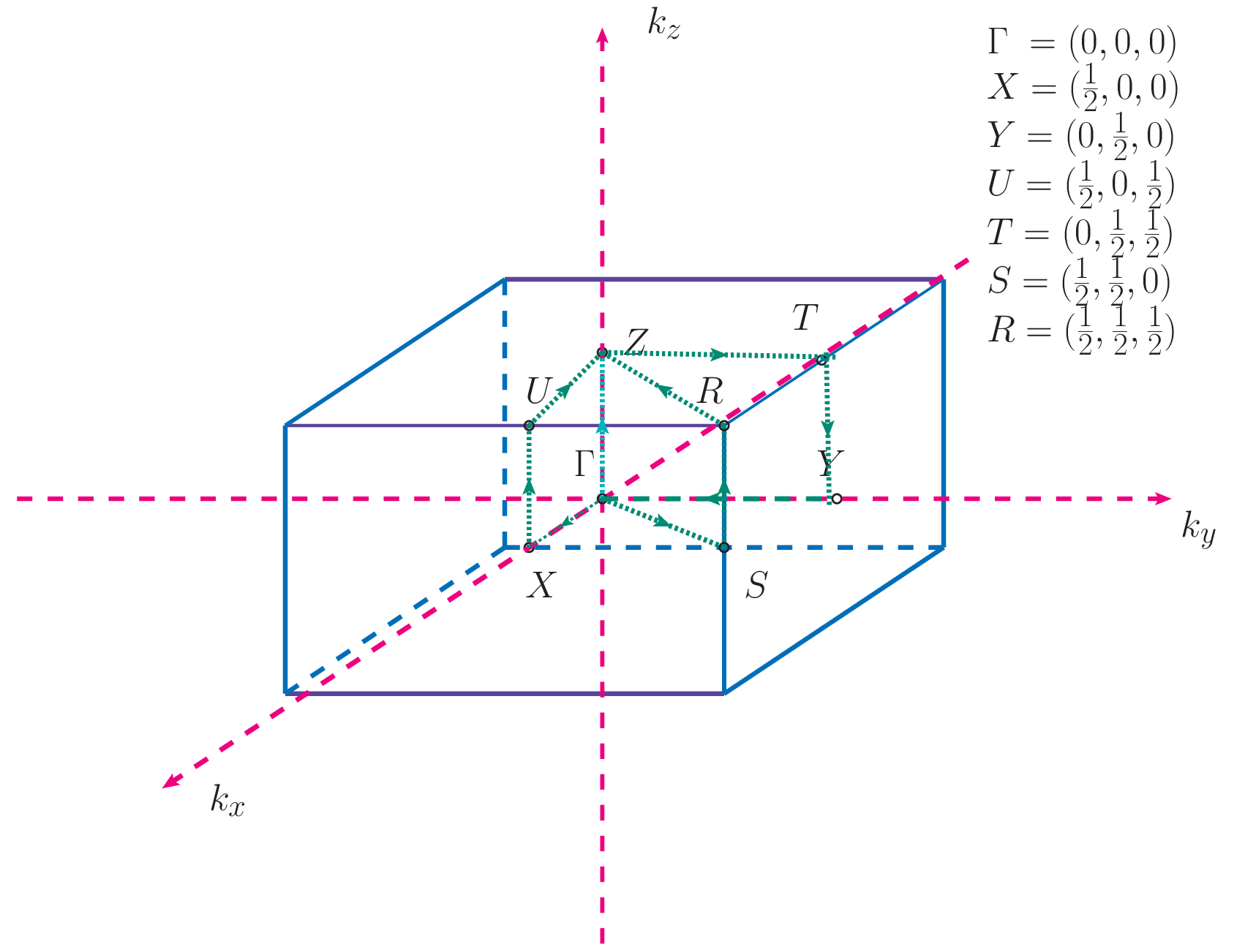} 
\caption{Left: Crystal Structure of (BA)$_2$PbI$_4$. Grey (purple) spheres denote the Pb (I) atoms. The gold spheres represent the C atoms  bonded to either each other or to small red (or bluish) hydrogen (or Nitrogen) atoms and, thus,
  form the  butylammonium  chains which intecalate the PbI$_4$ layers. 
  Middle: Halide-perovskite matrix. Right: The Brillouin zone and the path
  connecting the specific high-symmetry points used to compute the   band-structure.}
\label{c-structure}
\end{figure*}

Fig.~\ref{c-structure} illustrates the crystal structure of the low-temperature structure\cite{Menahem2021} of (BA)$_2$PbI$_4$
layered material. There are PbI$_4$ perovskite layers which are intercalated by butylammonium chains.
Notice that the atomic positions in two successive PbI$_4$ perovskite layers
are staggered relative to the atomic positions of the same atoms in their nearest neighboring layer. As a result the unit-cell of the Bravais lattice
is twice as long along the $c$ axis.

\begin{figure*}[!htb]
\includegraphics[width=0.3\linewidth]{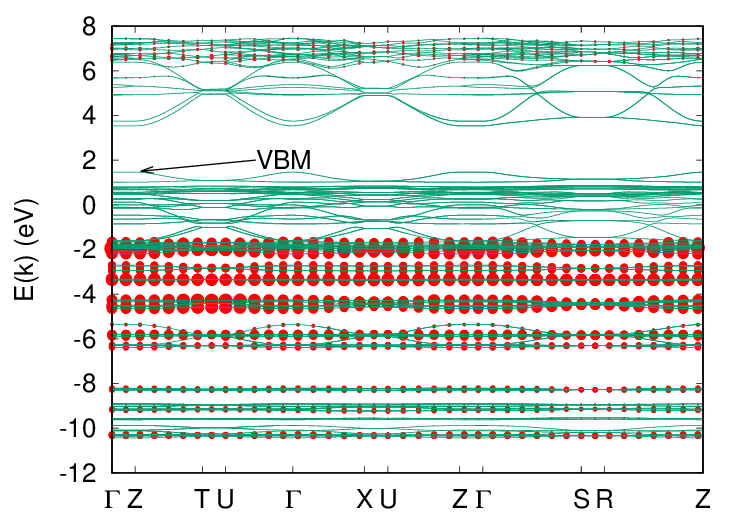} \hskip 0.1 in
\includegraphics[width=0.3\linewidth]{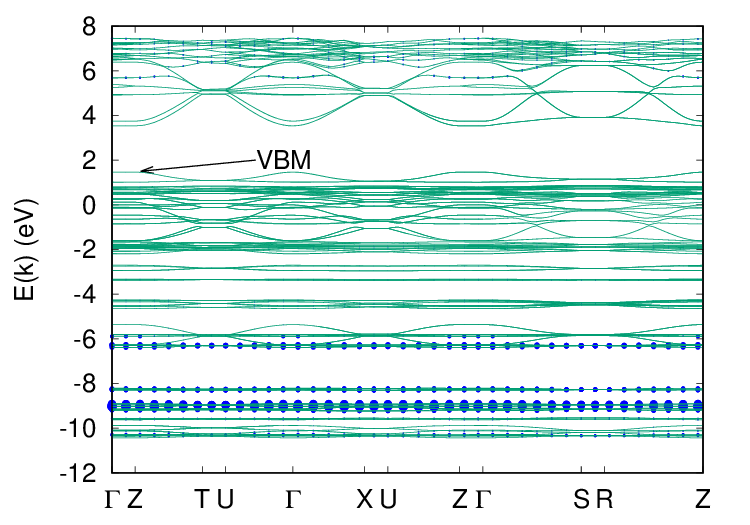} \hskip 0.1 in
\includegraphics[width=0.3\linewidth]{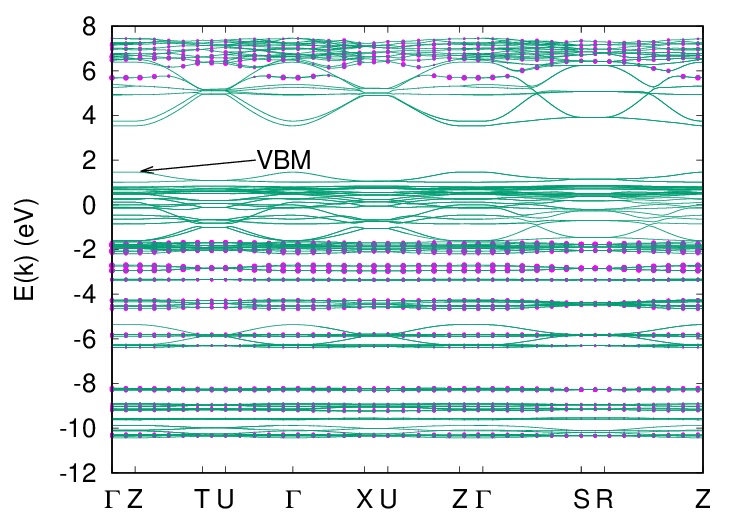} 
\caption{Projections of the Bloch states to all atomic orbitals of  C (left) N (middle) and H (right)
  present in the butylammonium ligands. All the bands are illustrated with both green lines and red (C)  blue (N) or magenta (H)
  symbols. The size of these symbols is proportional to the magnitude-squared
  of the projection of each band (for each $\vec k$ point) on these orbitals.
  As a result, because for many bands the projection is negligible, which gives
  rise to invisibly-small size of these symbols,
  only the lines are visible for these bands. Notice that there is no visible contribution
  of these orbitals to the bands within the 5 eV energy range of our focus,
  i.e., from 2 eV  above the CBM to 1 eV below the CBM.}
\label{Proj}
\end{figure*}

\subsection{(BA)$_2$PbI$_4$ Band Structure}

First, we carried out calculations for the (BA)$_2$PbI$_4$ structure\cite{PhysRevB.67.155405} shown in the left panel of
Fig.~\ref{c-structure} using the Quantum Espresso\cite{QE} (QE) implementation of the  
density functional theory (DFT) 
in the GGA framework.
The Perdew-Burke-Ernzerhof (PBE) exchange correlation  
functional\cite{Perdew1996_GGA_PBE} was
used with Projector Augmented-Wave \cite{PhysRevB.59.1758}
pseudopotentials generated with a scalar-relativistic calculation
local potential using
the ``atomic'' code by Dal Corso\cite{PhysRevB.82.075116}.
Our self-consistently converged ground-state calculation used a  $8\times 6\times 3$ k-point mesh and a 30 Rd energy cut-off.
In Appendix~\ref{convergence} a convergence study demonstrates that the k-point mesh and the energy cut-off used are accurate for the purpose of the present paper.

Our first objective is to establish that the atomic orbitals of the
butylammonium organic ligands, i.e., the orbitals of C, N and H, have negligible
contribution to the Bloch states with energy which falls in the energy
window of 1 eV below the top of the VBM and 2 eV above the
CBM, which, including the gap is about a 5 eV range. Fig.~\ref{Proj}
illustrates that there is insignificant projection of each of the
Bloch states to the local orbitals of all C, N and H.

\begin{figure}[!htb]
\includegraphics[width=1.0\linewidth]{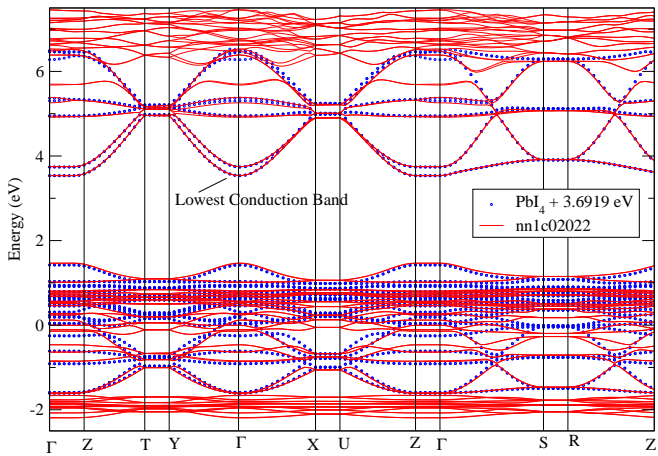} 
\caption{Comparison of the band-structure between the (BA)$_2$PbI$_4$ material and the PbI$_4$ matrix. Notice that we have shifted the band
  energy scale of PbI$_4$ by 3.6919 eV.}
\label{comparison}
\end{figure}

Using  the same
exchange correlation functional and pseudopotentials as in the
case of (BA)$_2$PbI$_4$, we carried out a self-consistent-field (SCF) DFT ground state calculation
for the ``bare-bone'' perovskite atomic matrix illustrated in the middle panel of Fig.~\ref{c-structure},
i.e., without the intercalating butylammonium  chains.
We then computed the bands along the same crystallographic directions
as for the (BA)$_2$PbI$_4$ for comparison. These bands are shown in
Fig.~\ref{comparison} as blue circles and are compared with those of the complete material
 (BA)$_2$PbI$_4$ (shown as red-lines).

Notice that the agreement between the bands near the Fermi level is very good.
The position of the Fermi level is different for the real material
(BA)$_2$PbI$_4$ as compared to the simple PbI$_4$ matrix because the
intercalating butylammonium  chains add more electrons to these bands,
thus, raising the Fermi level and making it a band insulator.
The important conclusion is that the bands near the Fermi level and their Bloch wavefunctions,
assuming the same atomic positions, are
determined by the PbI$_4$ matrix to a good degree of accuracy.

\subsection{Projecting Bloch states  near the Fermi energy to atomic orbitals}
\begin{figure*}[!htb]
\includegraphics[width=0.45\linewidth]{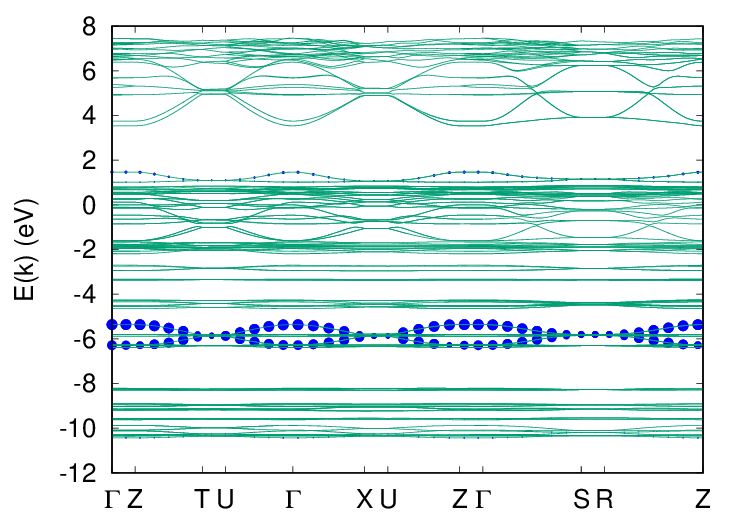} \hskip 0.1 in
\includegraphics[width=0.45\linewidth]{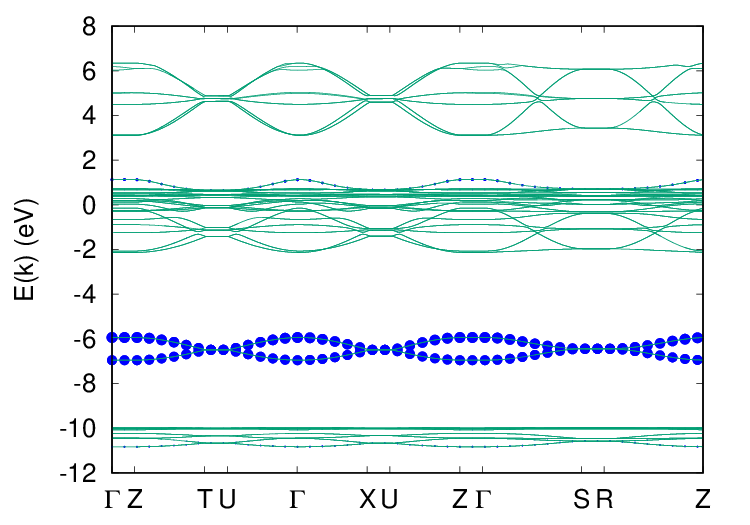} \hskip 0.1 in
\includegraphics[width=0.45\linewidth]{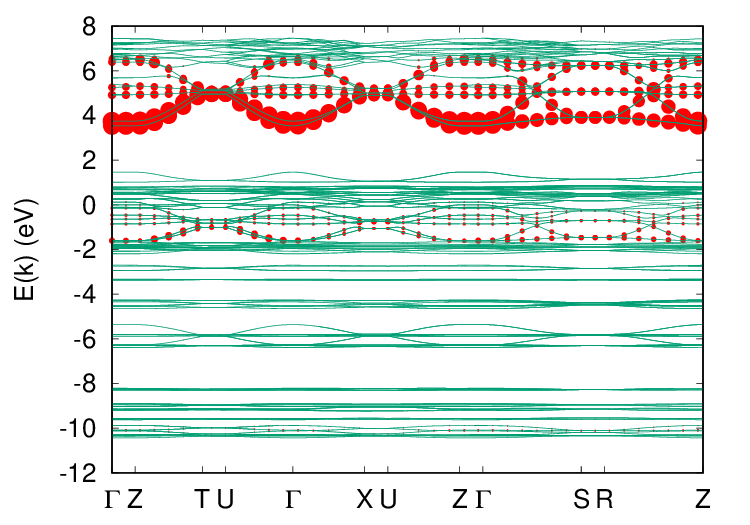} \hskip 0.1 in
\includegraphics[width=0.45\linewidth]{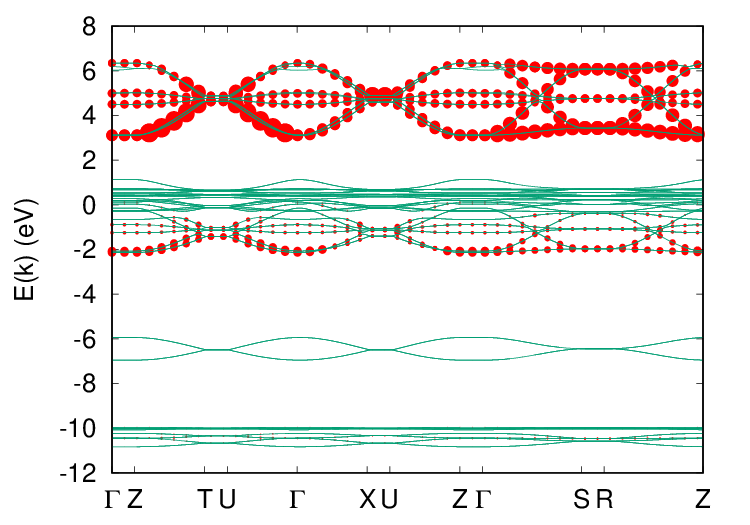} 
\caption{Projections of the Bloch states to local Pb $6s$ (top) and $6p$ (bottom) orbitals. 
  See caption of Fig.~\ref{Proj} for explanation of the symbol and lines in the plots.
  The panels on the left correspond to the (BA)$_2$PbI$_4$, while those on the right correspond to the PbI$_4$ halide-perovskite ``matrix''. The highest
occupied band is the one which starts at the $\Gamma$ point at just above 1 eV on this scale.}
\label{Pb-Proj}
\end{figure*}

\begin{figure*}[!htb]
\includegraphics[width=0.45\linewidth]{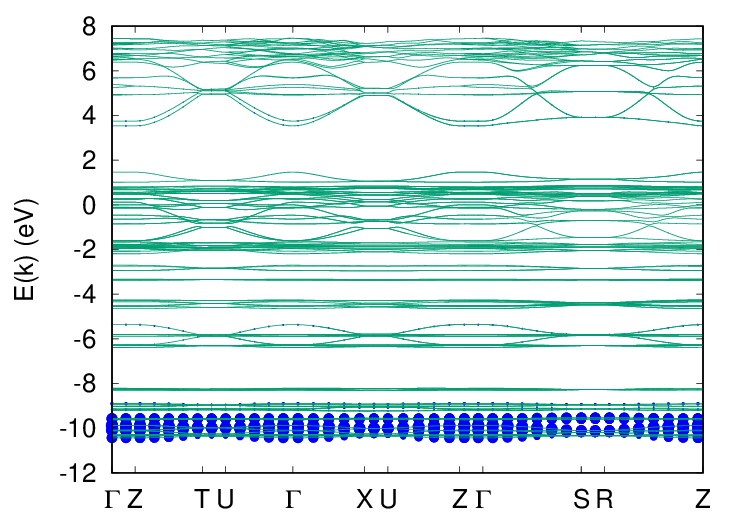} \hskip 0.1 in
\includegraphics[width=0.45\linewidth]{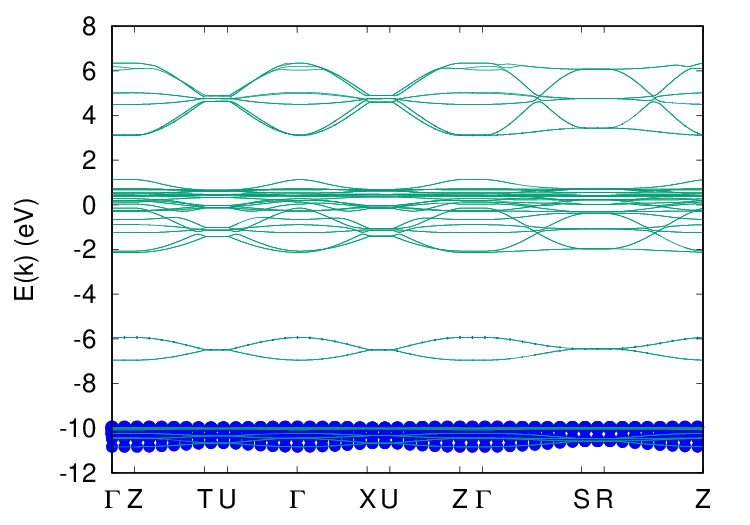} 
\includegraphics[width=0.45\linewidth]{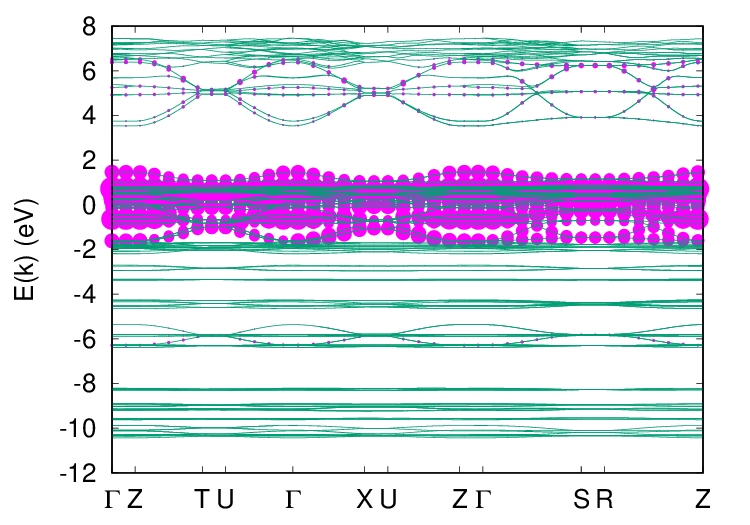} \hskip 0.1 in
\includegraphics[width=0.45\linewidth]{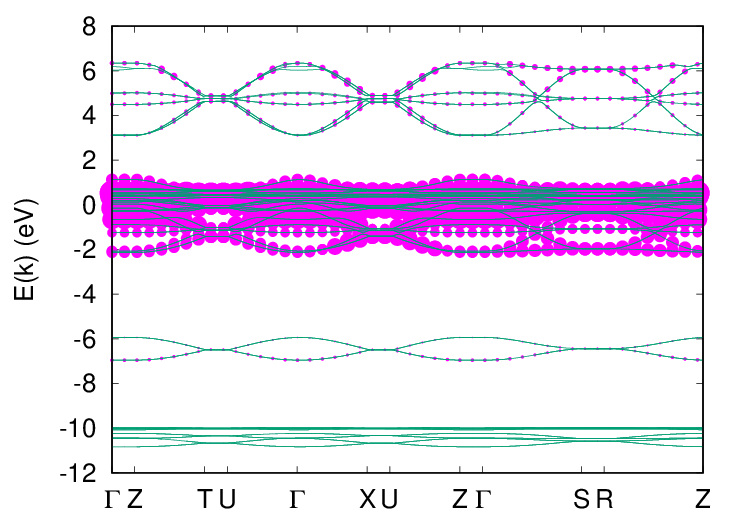} 
\caption{Projections of the Bloch states to local I $5s$ (top) and I $5p$ (bottom) orbitals. All the bands are illustrated with both lines and blue or magenta
  symbols.  See caption of Fig.~\ref{Proj} for explanation of the symbol and lines in the plots.
    The panels on the left row correspond to the (BA)$_2$PbI$_4$, while those on the right correspond to the PbI$_4$ halide-perovskite ``matrix''. The highest
occupied band is the one which starts at the $\Gamma$ point at just above 1 eV on this scale.}
\label{I-Proj}
\end{figure*}

Next, we wish to determine the orbital character of the bands
inside the energy window of our interest.
Since we have demonstrated in the previous section that these
bands are almost completely determined by the Pb and I atoms
we projected the bands in the orbitals of those atoms only.
In Figs.~\ref{Pb-Proj},\ref{I-Proj} the projection of
the bands in the atomic orbitals are shown. The size of the circle is
proposal to the contribution of the particular orbital to the given band.
The orbitals chosen are those that contribute to the bands and lie within $\pm 10$ eV from the Fermi level. These are the Pb outer orbitals, i.e.,
6$s$, $4f$ $5d$ and $6p$ and the I $5s$ $4d$ and $5p$.
Notice that the projections for the case of (BA)$_2$PbI$_4$ (left panels),
and those of the PbI$_4$ halide-perovskite ``matrix'' (right panels) are
very similar. In addition, we note that top valence bands and the lower conduction bands (i.e., with band energy less than about 5 eV and above 0 eV)
are mostly made out of $p$ (Pb or I) orbitals. There is
a very small amount of $s$ orbital contribution.

\subsection{Removing the organic molecules and adding Cs at the location of N sites: Cs$_2$PbI$_4$}

\begin{figure}[!htb]
\includegraphics[width=0.3\linewidth]{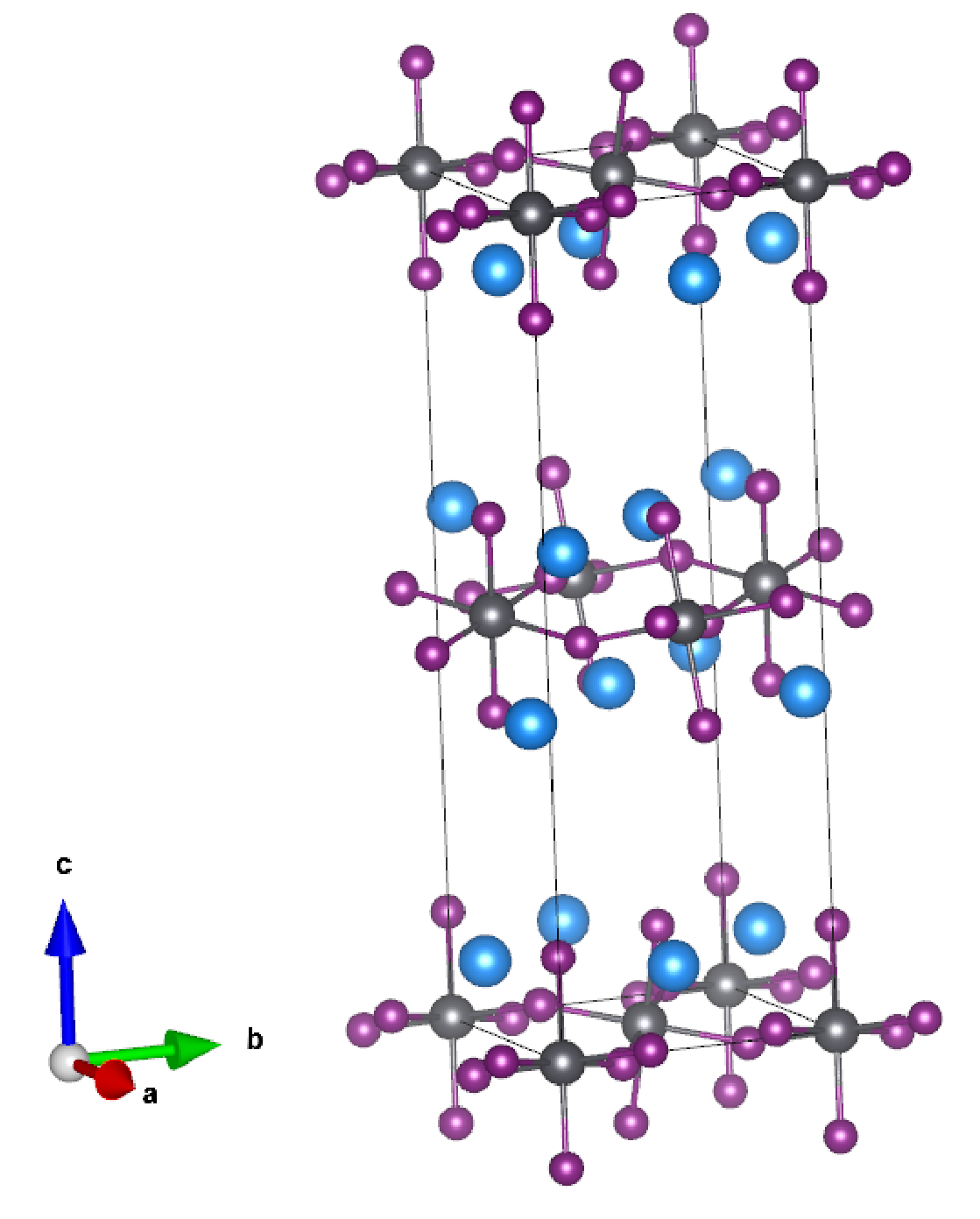} \hskip 0.3 in
\includegraphics[width=0.6\linewidth]{Fig7b.eps} 
\caption{Left: The structure of a compound obtained by removing all the
  butylammonium  chains and by adding Cs at the N sites of the original
  (BA)$_2$PbI$_4$
  compound.
  Right: We compare the bands
  of the PbI$_4$ (which are very close to those of (BA)$_2$PbI$_4$ as
  demonstrated in Fig.~\ref{comparison}) with  the bands of Cs$_2$PbI$_4$ obtained by a DFT calculation for this compound, shifting the energy of all the bands of PbI$_4$ by a constant amount of 2.4749 eV.}
\label{cs-compound}
\end{figure}

In order to clearly demonstrate the almost irrelevance of the
butylammonium  chains on the band-structure near the Fermi level, we calculated
the band-structure of the a compound obtained by removing all the butylammonium  chains and by adding Cs at the N sites of the original (BA)$_2$PbI$_4$
  compound. We choose Cs based on its electronegativity relative to the
  Fermi level of the PbI$_4$ matrix, in order to make sure that the energy level of the outer $s$
  Cs level falls above the energy of the highest occupied band, such that these Cs $s$ states will be empty inside
  the compound, thus, resulting in doping the PbI$_4$ matrix in the same manner
  as the alkylammonium  chains cause doping.
  The structure is illustrated in the left panel of Fig~\ref{cs-compound}
  and the DFT calculated band structure is compared to the (BA)$_2$PbI$_4$
  compound in right panel of Fig.~\ref{cs-compound}.

  This compound exists
  in nature but
  with different lattice constants than those used in this calculation:
  we wish   to keep them the same as in the (BA)$_2$PbI$_4$ for direct
  comparison of the bands.
  Notice that after shifting the energy of all the bands of PbI$_4$ by the same amount of 2.4749 eV
  there is very good agreement between the bands near the Fermi energy.
  This finding strengthens our conclusion and makes transparent our
  statement that the bands of the PbI$_4$ matrix  almost solely determine the band structure
 within 1 eV below the VBM  and 2 eV above the bottom of the CBM, which, including the energy gap, is about a 5 eV energy range.

   \section{Tight-binding model}
   \label{sec:TB}

From the previous comparison we conclude that it makes sense to derive
a TB model to
describe these rigid-band features. In this model the role of all the interlayer butylammonium  chains is to create the structure, and their role in the
electronic structure is to provide one additional electron per Pb atom.
Therefore, in our TB model which aims at describing these
rigid band features for only the bands in the energy window of our interest
discussed previously, the intercalating organic chains will be ignored.

We start from  a symmetric
system without any octahedra distortions and we choose the $a$-axis and $b$ axis to be of the same length, i.e., the $a-b$ lattice has the symmetry of the
square-lattice. In the real material the $a$ and $b$ axes are very close
to each other, namely, $|\vec b|/|\vec a| \sim 0.98 $.

First, we ignore the interplane coupling, which reduces the effective unit-cell
to one with half the number of atoms. Notice that the bands obtained by the
DFT calculation have different bandwidths along the $c$ axis.
The bands that are mostly made of Hydrogen atomic orbitals, i.e., from the
butylammonium chains have significant bandwidth of the order of 0.1 eV.
The bands of our interest, however, i.e., those within the region of
5 eV near the Fermi level, have remarkably negligible dispersion along the $c$ axis. The bandwidth of these bands is less than 0.01 eV. This justifies
our treatment of these bands as two-dimensional.

\begin{figure}[!htb]
\includegraphics[width=0.45\linewidth]{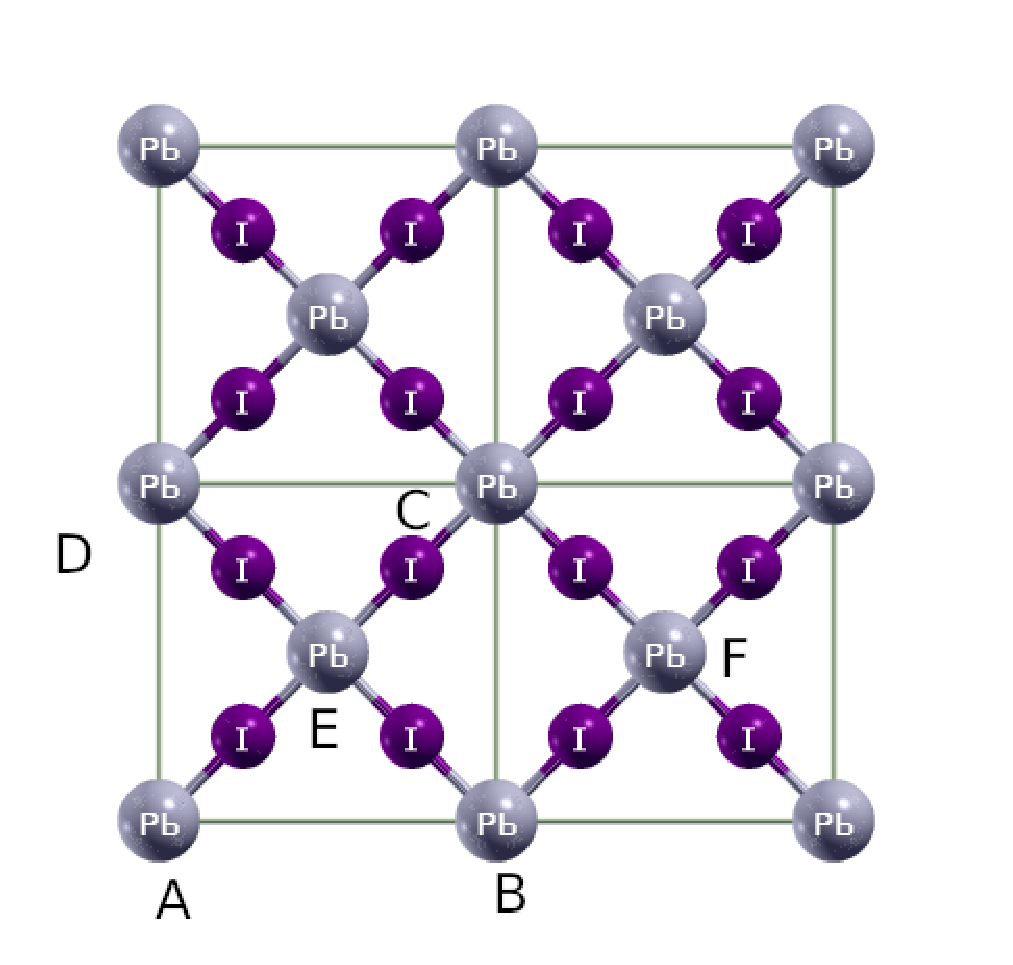} \hskip 0.1 in 
\includegraphics[width=0.45\linewidth]{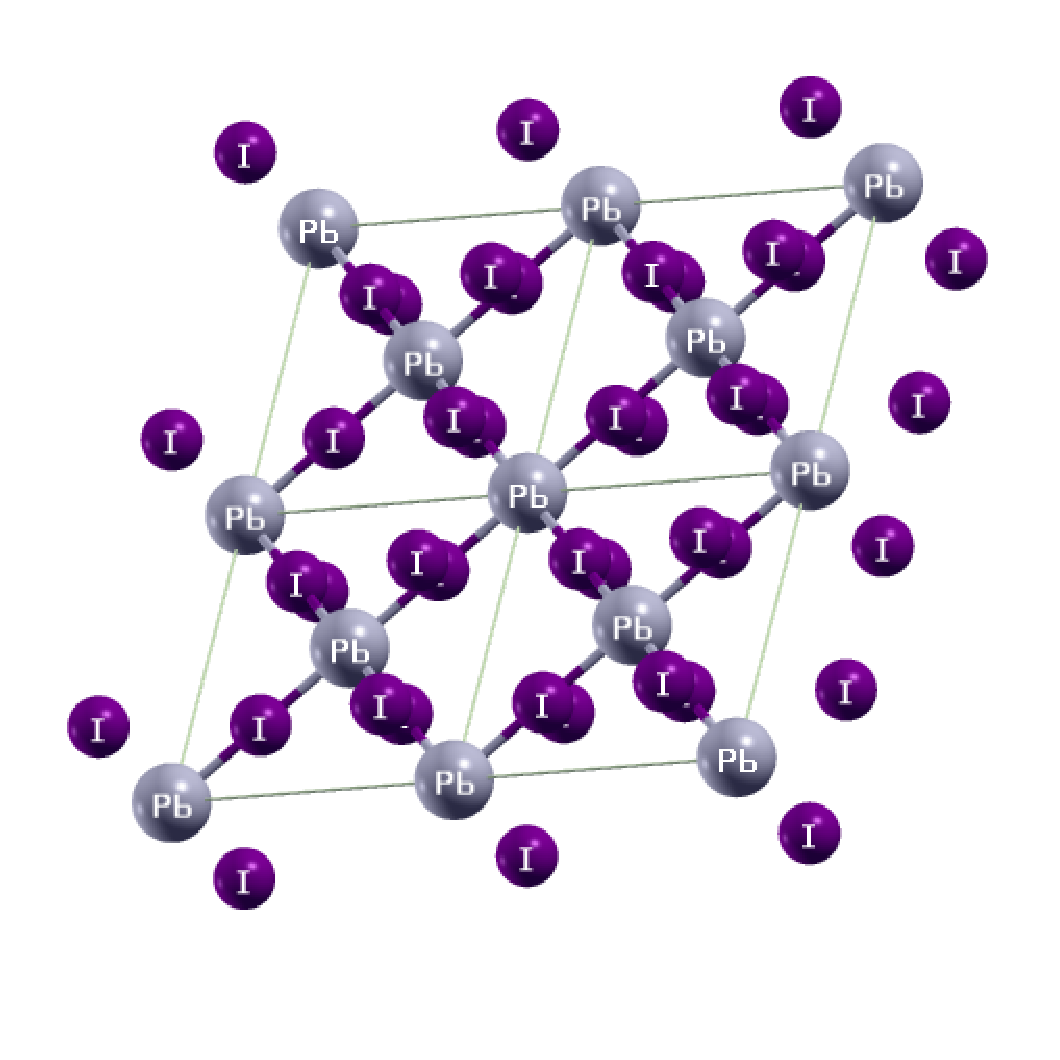} 
\caption{Left: The pure two-dimensional structure. Right: A single layer with all the Pb and I atoms but with un-distorted octahedra. }
  \label{reduced-structure}
\end{figure}

In addition, as a first step, we are going to ignore the role of the I atoms which are off the plane which reduces the problem to the 2D unit-cell shown on the left panel of Fig.~\ref{reduced-structure}.  We will also discuss how to include these off-planar atoms displayed on the right panel of Fig.~\ref{reduced-structure} as a second stage.

Next, we are going to consider matrix elements of the Hamiltonian between
the following twelve (12) states inside the unit-cell of the reduced
Bravais lattice: $|s^{(i)}\rangle$ $| p^{(i)}_x \rangle$,
$| p^{(i)}_y \rangle$ and $| p^{(i)}_z \rangle$, where $i=1,2,3$ for the 3 atoms
in the reduced unit cell, i.e., the Pb atom at the origin $(0,0,0)$, the
I atom at $(a/2,/a/2,0)$ and the I atom at $(-a/2,a/2,0)$.
We note that throughout the rest of the paper,
our $x$ and $y$ axes are with respect to the $45^{\circ}$ rotated
coordinate system, not the original unit cell.
The state $|s^{(i)}\rangle$ for $i=1$ is the
Pb $6s$ orbital and for $i=2,3$ it is the $5s$ I orbital.
The states $|p^{(i)}_x\rangle$ and $|p^{(i)}_y\rangle$
where $i=1$ corresponds to the Pb $6p$ orbitals, whereas when $i=2,3$
it corresponds to the I $5p$ orbitals.

Matrix elements of the form $\langle s^{(i)} | \hat H | s^{(j)} \rangle$
when $i=j$ are the on-site energies, which are $E^{(1)}_s$, $E^{(2)}_s=E^{(3)}_s$, while off-diagonal matrix elements are non-zero only if the atoms are nearest neighbors, in which case they are $V_{ss} \equiv \langle s^{(1)} | \hat H | s^{(2)} \rangle = \langle s^{(1)} | \hat H | s^{(3)} \rangle$.

\begin{figure}[!htb]
\includegraphics[width=0.45\linewidth]{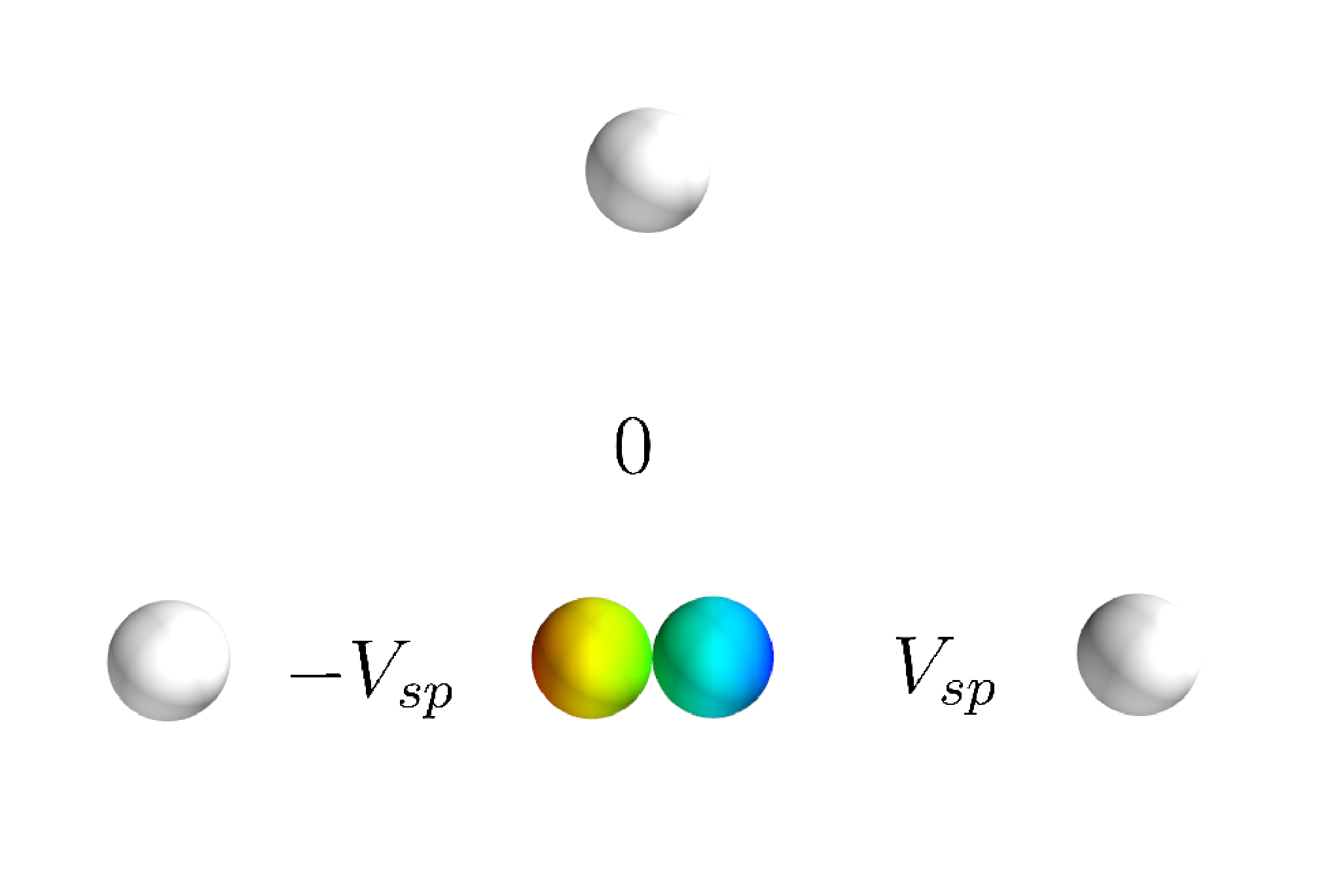}\hskip 0.1 in 
\includegraphics[width=0.45\linewidth]{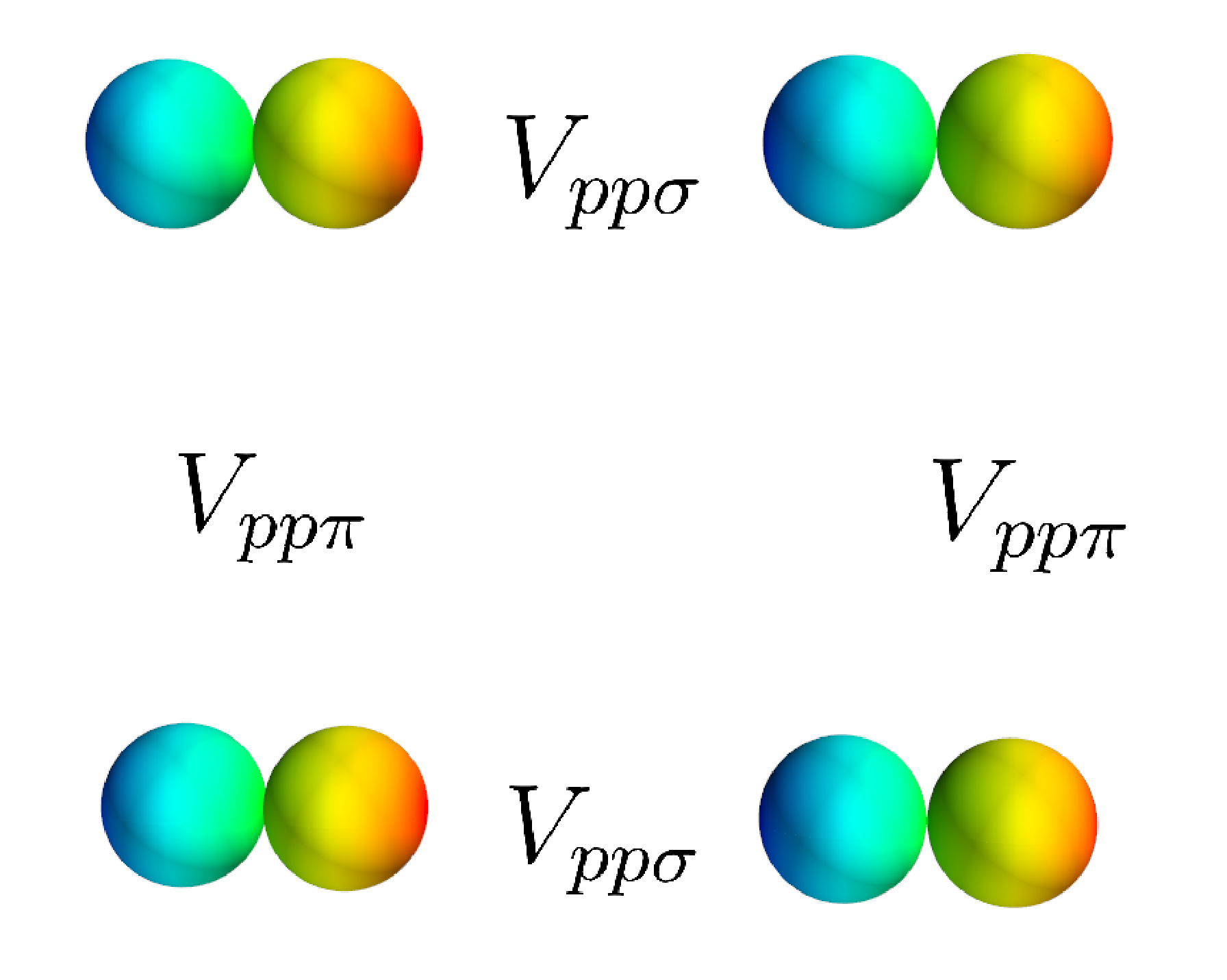} 
\caption{Left: Matrix elements of $\hat H$ between a $p_x$ orbital and an $s$-type orbital. The green (yellow) regions of the $p$ orbitals denote negative (positive) sign. As a result these matrix
  elements can be zero or they can have opposite sign. Right: The $\sigma$- and $\pi$-type matrix elements between $p$ orbitals of
neighboring atoms.  $\sigma$  ($V_{pp\sigma}$) bonds form when the orbital orientation is along the line connecting the atoms, while $\pi$  ($V_{pp\pi}$) bonds form when the orbital orientation is perpendicular to the line connecting the atoms. When two orbitals have perpendicular orientations they do not yield a non-zero matrix element.}
\label{bonds}
\end{figure}

Matrix elements between a $6s$ orbital of a Pb atom and a $5p$-type orbital
of any of its nearest neighbor I atoms or vice-versa, when non-zero, are given by $V_{sp}$ or $V_{ps}$. Because of the negative eigenvalue of the $p_x$ orbital with respect to reflections about a plane perpendicular
to the $x$-axis, some of these matrix elements are zero and others have a relative minus sign as illustrated in the left panel of Fig.~\ref{bonds}.

There two kinds of matrix elements of $H$ between the same type of $p$-orbitals as illustrated in the right panel of Fig.~\ref{bonds}. As explained in the figure caption of this figure,
they can be either of $\sigma$-type, i.e., $V_{pp\sigma}$ or of $\pi$-type, i.e., $V_{pp\pi}$. One has to be careful about their relative sign and when these matrix elements are zero.

\begin{figure}[!htb]
\includegraphics[width=0.8\linewidth]{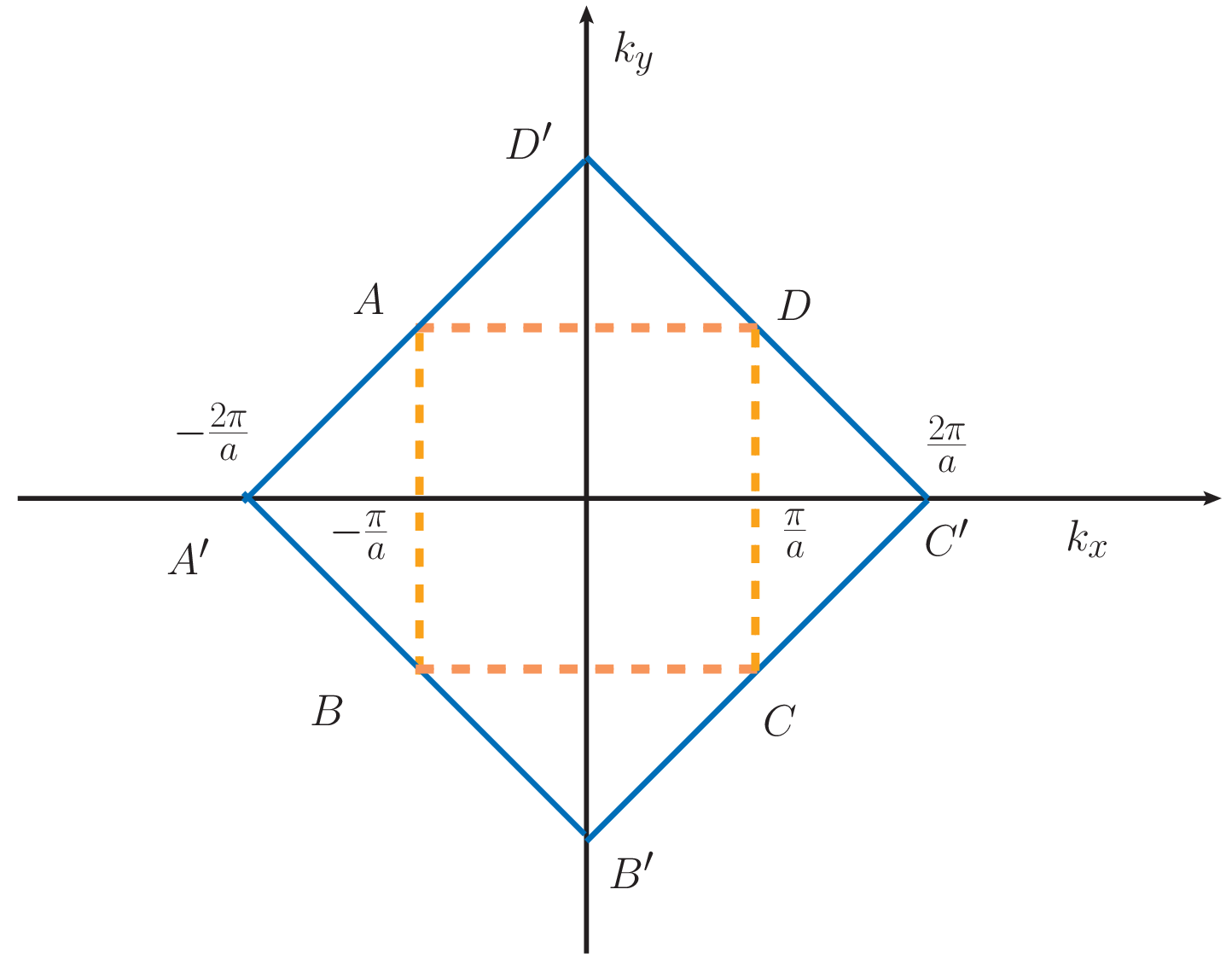} 
\caption{The Brillouin zones of the original and the reduced simpler structure.}
  \label{BZ}
\end{figure}

\begin{table*}[!htb]
\scriptsize
		\begin{tabular}{ |c|c|c|c|c|c|c|c|c|c|c|c|c|}
			\hline
  & $|s^{(1)}\rangle$ & $|p_x^{(1)}\rangle$ & $|p_y^{(1)}\rangle$ & $|p_z^{(1)}\rangle$ & $|s^{(2)}\rangle$ & $|p_x^{(2)}\rangle$ & $|p_y^{(2)}\rangle$ & $|p_z^{(2)}\rangle$ & $|s^{(3)}\rangle$ & $|p_x^{(3)}\rangle$ & $|p_y^{(3)}\rangle$ & $|p_z^{(3)}\rangle$ \\

			\hline

$\langle s^{(1)} |$ & $E^{(1)}_{s}$ & 0 & 0 & 0 & $V_{ss} d_{x}$ & $V_{sp} d^{\prime}_x$ & 0 & 0 & $V_{ss} d_{y}$ & 0  & $V_{sp} d^{\prime}_y$ & 0 \\

			\hline
$\langle p_x^{(1)} |$ & 0 & $E^{(1)}_{p_{xy}}$  & 0 & 0 & $-V_{ps} d^{\prime}_x$ &$V_{pp\sigma}d_x$ & 0 & 0 &0&$V_{pp\pi}d_y$  &0 &0\\

			\hline
$\langle p_y^{(1)} |$ & 0 & 0 & $E^{(1)}_{p_{xy}}$& 0 & 0 &0& $V_{pp\pi}d_x$ &0 & $-V_{ps}d^{\prime}_y$ &0 & $V_{pp\sigma} d_y$& 0 \\

			\hline
$\langle p_z^{(1)} |$ & 0 & 0 & 0 & $E^{(1)}_{p_z}$&0 &0 & 0&$V_{pp\pi} d_x$ & 0 &0 & 0  & $V_{pp\pi} d_y$\\

			\hline
$\langle s^{(2)}| $& $V_{ss}d_x$ & $V_{ps}d^{\prime}_x$ & 0 &0 & $E^{(2)}_{s}$ & 0& 0 &0 &0 &0 & 0& 0   \\

			\hline
$\langle p_x^{(2)} |$ &-$V_{sp}d^{\prime}_x$ & $V_{pp\sigma} d_x$&0 & 0&0 & $E^{(2)}_{p_{xy}}$ &0 &0 &0 &0 & 0  &0\\

			\hline
$\langle p_y^{(2)} |$ &0 &0 &$V_{pp\pi}d_x$ & 0 &0 &0 & $E^{(2)}_{p_{xy}}$&0 &0 & 0 & 0 &0\\

			\hline
                        $\langle p_z^{(2)} |$ &0 &0 &0 &$V_{pp\pi}d_x$ & 0 &0 &0 &
                        $E^{(2)}_{p_{xy}}$&0 &0 & 0 & 0\\
                        
			\hline
$\langle s^{(3)} |$ &$V_{ss}d_y$ &0 &$V_{ps}d^{\prime}_y$ &0 &0 &0 &0 &0 &$E^{(2)}_{s}$ &0 &0 &0 \\

			\hline
$\langle p_x^{(3)} |$ & 0 &$V_{pp\pi} d_y$ &0 &0 &0 &0 &0 &0 &0 &$E^{(2)}_{p_{xy}}$ &0 &0 \\

			\hline
$\langle p_y^{(3)} |$ &$-V_{sp}d^{\prime}_y$ & 0& $V_{pp\sigma}d_y$&0 &0 &0 &0 &0 &0 &0 &$E^{(2)}_{p_{xy}}$ & 0\\

			\hline
$\langle p_z^{(3)} |$ &0 &0 &0 &$V_{pp\pi}d_y$ &0 &0 &0 &0 &0 &0 &0 & $E^{(2)}_{p_z}$\\

			\hline
		\end{tabular}
		\caption{Tight-binding matrix for the truncated 2D case. }
		\label{table:1}

\end{table*}
\addtolength{\oddsidemargin}{0in}

\begin{table}[!htb]

  \begin{tabular}{ |c|c|c|c|c|c|c|c|c|c|c|}
    \hline
    $E^{(1)}_s$ & $E^{(1)}_{p_{xy}}$ & $E^{(1)}_{p_{z}}$ &
    $E^{(2)}_s$ & $E^{(2)}_{p_{xy}}$ & $E^{(2)}_{p_{z}}$  & $V_{pp\sigma}$ & $V_{pp\pi}$ &
    $V_{sp}$ & $V_{ss}$ & $V_{ps}$ \\
    \hline
    -9.3 & -0.2 & -0.2 & -13.05 & -2.8 & -3.0 &  -2.5 & 0.8 & 1.2 & -0.6 & 0.8 \\
\hline    
  \end{tabular}
  \caption{Tight-binding parameters after fitting the DFT bands of the x-y symmetric
    un-tilted PbI$_4$ matrix. The fit is shown in Fig.~\ref{s-pbi4-bands}. }
		\label{table:2}

\end{table}
First, instead of using the unit-cell illustrated in
the left panel of Fig.~\ref{reduced-structure} by the ABCD square, which is the
one that is necessary to use when the symmetry is broken by the octahedra
tilting, we use as unit-cell the smaller size square EBFC which is rotated
by $45^{\circ}$ with respect to the original. This unit-cell contains only
one Pb and two I atoms and, therefore, 12 states. This doubles the size
of the BZ from the orange $ABCD$ square of Fig.~\ref{BZ} to the square labeled
$A^{\prime}B^{\prime}C^{\prime}D^{\prime}$, which is rotated by $45^{\circ}$ with respect to the 
x-axis. Therefore, for our convenience, we will solve the problem in this larger BZ (where we have only 12 states for each $k$) and in order to
compare with the results of the original BZ, we will need to fold this BZ to the
smaller size one.
This folding will increase the number of bands by a factor of two
and, therefore, we will be able to recover the total number of 24 bands,
which were present in the original unit-cell of the Bravais lattice.

If we include the matrix elements illustrated in Fig.~\ref{bonds}
and we transform our Hamiltonian in momentum space,
the Hamiltonian matrix becomes momentum-diagonal i.e.,
\begin{eqnarray}
  \hat H = \sum_{\vec k \in BZ} {\tilde H}_{\vec k},
\end{eqnarray}
where the sum is over the entire Brillouin zone (blue square of Fig~\ref{BZ})
and ${\tilde H}_{\vec k}$ is
a $12 \times 12$ matrix given in Table~\ref{table:1}, where
\begin{eqnarray}
d_x &=& 2 \cos({{(k_x+k_y)a} \over 4}), d_y = 2 \cos({{(k_x-k_y)a}\over 4}), \\
d^{\prime}_x &=& 2 i\sin({{(k_x+k_y)a} \over 4}),
d^{\prime}_y = 2 i\sin({{(k_x-k_y)a} \over 4}).
\end{eqnarray}
The subscripts $x$ and $y$ in the above $\vec k$-dependent
coefficients are labeled according to 
our $x$ and $y$ axes  which are with respect to the $45^{\circ}$ rotated
coordinate system, not the original unit cell. The $k_x$ and $k_y$, however, are
the $x$ and $y$ projections of $\vec k$ on the corresponding $x$ and $y$ axes
of the original un-rotated unit-cell. We use the latter for making
comparison with the bands obtained by DFT.
\begin{figure}[!htb]
\includegraphics[width=1.0\linewidth]{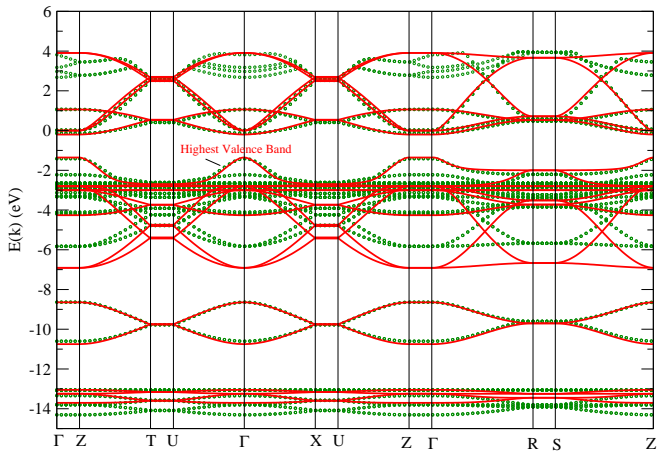} 
\caption{Comparison of the results obtained by TB with those by DFT
  for the symmetric PbI$_4$ matrix.}
\label{s-pbi4-bands}
\end{figure}
A numerical diagonalization of the above $12 \times 12$ at every $\vec k$-point
leads to the results illustrated in Fig.~\ref{s-pbi4-bands} which are
compared to the results of the DFT of a 3D PbI$_4$ crystal without octahedral distortions and with $|\vec a| = |\vec b|$. The results of the fitting parameters
is given in Table~\ref{table:2}. We note that the disagreement for the highest energy bands is due to the fact that these bands are at positive (unbound)
energy and, therefore, the DFT calculation has included the effects of unbound electronic states. The DFT calculation involves more states because it includes the off-plane I atoms (see structure illustrated in the right panel of Fig.~\ref{reduced-structure}). The
TB matrix (a $20\times 20$ matrix) which includes the
$s$ and $p$ orbital of these atoms is given in Appendix~\ref{tight-binding20}. There are
several more fitting parameters to use in this case and the agreement
can be improved by introducing some of the missing bands away from the Fermi level. Such an approach, however,
would lead to significant complication which works against our goal
of simplifying the problem and leaving the description near the Fermi energy
as accurate as possible. We have implemented this more complex TB
Hamiltonian but the fact that it yields 16 additional bands (when we fold the BZ) leads to a
fitting procedure which does not have a unique and simple solution.
For completeness, however, we provide this more complex TB Hamiltonian matrix in the Table of Appendix~\ref{tight-binding20} and we leave out
from this paper the ambiguous results of such a fit.

In Sec.~\ref{full-crystal-fit} we will use our TB model to fit the results
of the actual 3D (BA)$_2$PbI$_4$ crystal, where we find that the quality of the fit is equally good.
First, however, we would like to discuss an analytical treatment of the problem, which we do in the next Section.

\section{Analytical description}
\label{analytical}
Here, we describe a simplified analytical treatment of the problem which 
contains all the essential elements of the original system near the Fermi
level.
\subsection{Non-interacting $s$ and $p$ orbitals}
First, we define the model subspace, which is all the orbitals that fall inside the energy range of our interest
which is 1 eV below the VBM and 2 eV above the CBM. These states are the Pb $6p$ and the I $5s$ orbitals.
The Hamiltonian acting inside this model subspace is the following:
\begin{eqnarray}
  \hat H_M = \hat H_s + \hat H_{p_x} + \hat H_{p_y} + \hat H_{p_z},
\end{eqnarray}
where $\hat H_s$ corresponds to the orbital $|s^{(1)}\rangle$, i.e., the Pb $6s$ orbital, i.e.,
\begin{eqnarray}
 \hat H_s = E^{(1)}_s | s^{(1)} \rangle \langle s^{(1)} |,
\end{eqnarray}
and
\begin{eqnarray}
\hat H_{p_{\alpha}} &=& \left ( \begin{tabular}{ccc} 
$E^{(1)}_{p_{\alpha}}$ & $V^{(\alpha)}_1$& $V^{(\alpha)}_2$  \\
$V^{(\alpha)*}_1$ & $E^{(2)}_{p_{\alpha}}$ & 0   \\
  $V^{(\alpha)*}_2$ & 0 & $E^{(2)}_{p_{\alpha}}$  \end{tabular} \right )
\end{eqnarray}
in the basis $|p^{(1)}_\alpha\rangle,|p^{(2)}_\alpha\rangle,|p^{(3)}_\alpha\rangle$
where 1,2,3 stand for the Pb and the two $p_\alpha$ ($\alpha=x,y,z$) orbitals of its neighboring I atoms. Here
\begin{eqnarray}
  E^{(1,2)}_{p_{x}}&=& E^{(1,2)}_{p_{xy}}, V^{(x)}_1 =  d_x V_{pp\sigma},
 V^{(x)}_2 =  d_y V_{pp\pi}, \\
E^{(1,2)}_{p_{y}}&=& E^{(1,2)}_{p_{xy}}, V^{(y)}_1 =  d_x V_{pp\pi},
 V^{(y)}_2 =  d_y V_{pp\sigma}, \\
E^{(1,2)}_{p_{z}}&=& E^{(1,2)}_{p_{z}}, V^{(z)}_1 =  d^{\prime}_x V_{pp\pi},
V^{(z)}_2 =  d^{\prime}_y V_{pp\pi}
\end{eqnarray}
Next, we will include the coupling of this subspace to the states which are outside this energy range
but not too far away. For example, we will include the coupling between the Pb $6s$ and the I $5p$ because the latter falls inside the
subspace of our interest, but we will ignore the
coupling $V_{ps}$ between Pb $6p$ and the I $6s$ state as the latter state is too
far below ($~13$ eV) the VBM and  $V_{ps}$ is only 0.8 eV, namely, $V_{ps}/(E^{(1)}_p - E^{(2)}_s) \sim 0.06$.
We will consider $V_{ss}=0$ as this couples the Pb $6s$ and I $5s$ orbitals which are far from the above energy range and
both states fall outside the energy domain of our interest. 
The resulting Hamiltonian is given by
\begin{eqnarray}
  \hat H = \hat H_M + \hat H_{sp},
\end{eqnarray}
The $\hat H_{sp}$ couples the Pb $s$ (i.e., $|s^{(1)} \rangle$) with
$|p^{(2)}_x \rangle$ and $|p^{(3)}_y \rangle$.
\begin{eqnarray}
\hat H_{sp} &=& \left ( \begin{tabular}{ccc} 
$E^{(1)}_{s}$ & $V_{sp} d^{\prime}_x$& $V_{sp} d^{\prime}_y$  \\
$-V_{sp} d^{\prime}_x$ & $E^{(2)}_{p_{xy}}$ & 0   \\
  $-V_{sp} d^{\prime}_y$ & 0 & $E^{(2)}_{p_{xy}}$  \end{tabular} \right ),
\label{hsp}
\end{eqnarray}
Diagonalization of the $3\times 3$  matrices $\hat H_{p_x}$, $\hat H_{p_y}$, $\hat H_{p_z}$, of the model subspace yields the following eigenvalues:
\begin{eqnarray}
  E^{\alpha}_{\pm}(\vec k) &=& {{E^{(1)}_{p_{\alpha}} + E^{(2)}_{p_{\alpha}}} \over 2}
  \pm R_{\alpha} \hskip 0.2 in E^{\alpha}_0 (\vec k) = E^{(2)}_{p_{\alpha}},  \\
  R_{\alpha} &\equiv& \sqrt{\Bigl ({{E^{(1)}_{p_{\alpha}} - E^{(2)}_{p_{\alpha}}} \over 2}\Bigr)^2 +
    |V^{(\alpha)}_1|^2 + |V^{(\alpha)}_2|^2}.
  \label{eigenvalues}
\end{eqnarray}
The corresponding eigenstates for each of the above cases are given by the following form:
\begin{eqnarray}
  |\psi^{(\alpha)}_{\pm} \rangle &=& {1   \over {\xi^{\alpha}_{\pm}}}
\Bigl [(E^{(2)}_{p_{\alpha}} - E^{(\alpha)}_{\pm}) | p^{(1)}_{\alpha}  \rangle \nonumber \\ 
      &+&    V^{(\alpha)}_1 | p^{(2)}_{\alpha} \rangle + V^{(\alpha)}_2 | p^{(3)}_{\alpha} \rangle \Bigr ]  , \\
  | \psi^{(\alpha)}_0 \rangle &=& 
  {{ V^{(\alpha)}_2 | p^{(2)}_{\alpha} \rangle - V^{(\alpha)}_1 | p^{(3)}_{\alpha} \rangle }  \over {\eta^{(\alpha)}}}, \\
  \xi^{(\alpha)}_{\pm} &=& \sqrt{ (E^{(2)}_{p_{\alpha}} - E^{\alpha}_{\pm})^2 + |V^{(\alpha)}_1|^2 + |V^{(\alpha)}_2|^2},\\
  \eta^{\alpha)} &\equiv&\sqrt{|V^{(\alpha)}_1|^2 + |V^{(\alpha)}_2|^2}.
  \label{eigenstates}
\end{eqnarray}
\begin{figure}[!htb]
\includegraphics[width=0.8\linewidth]{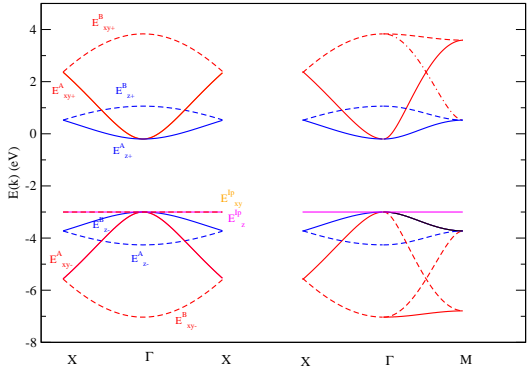} 
\caption{The bands of the simple analytical 2D tight-binding model along some high symmetry directions without including the
  coupling between Pb $6s$ and I $5p$.}
\label{simple-bands}
\end{figure}
\begin{figure}[!htb]
\includegraphics[width=0.8\linewidth]{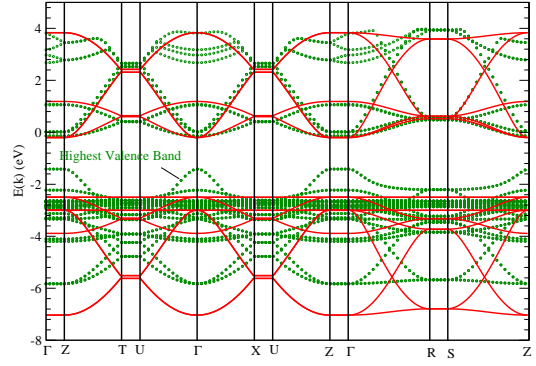} 
\caption{Comparison with the DFT results of the 2D PbI$_4$ crystal with the results of our analytical 2D tight-binding model without including the
  coupling between Pb $6s$ and I $5p$. }
\label{model-no-sp}
\end{figure}
The unit-cell doubling and $45^{\circ}$ rotation leads to the folding of the
BZ which doubles the number of eigenvalues to 18 (some of which are degenerate
in this higher symmetry Hamiltonian). They are illustrated in
Fig.~\ref{simple-bands} and are compared to the DFT calculation in Fig.~\ref{model-no-sp} using the same values of the parameters of Table~\ref{table:2}
but ignoring the $s$ orbitals completely, i.e.,  $V_{ss}=V_{sp}=V_{ps}=0$.
Notice that the occupied bands are described well, notice, however, 
that the top of the valence band (i.e., the I $5p$ state) is flat and the model does not describe
its dispersion. The main reason for this discrepancy is its coupling to the Pb $6s$
and it is corrected in the next subsection.

\subsection{Integrating out the Pb $s$ orbitals}

The I $5p$ bands $E^{0}_x$ or $E^{0}_{y}$ are flat in this approximation.
Next we  include the role of the Pb $6s$ and I $5p$ hybridization which will
account for the dispersion of the top valence band (which corresponds
to the flat band illustrated by the green-line in Fig.~\ref{simple-bands}.
The hybridization of the Pb $6s$ orbital with the I $5p$, i.e.,
the matrix element $V_{sp}$, couples the $| s^{(1)} \rangle$
and the $| p^{(2)}_x\rangle$ and $| p^{(2)}_y \rangle$ as in Eq.~\ref{hsp}.

\begin{figure}[!htb]
\includegraphics[width=0.8\linewidth]{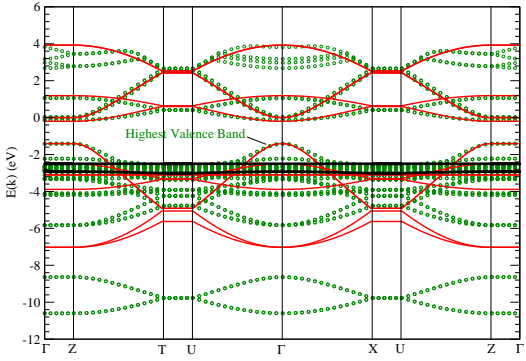} 
\caption{Comparison of the bands with the DFT results after
  including the role of the $V_{sp}$ interaction in our simple model.}
\label{model-with-sp}
\end{figure}

This interaction is crucial when there is an exact or almost degeneracy
at specific $\vec k$
points between the  $E^{\alpha}_{\pm}(\vec k)$ and $E^{\alpha}_0(\vec k)$ bands
given by Eq.~\ref{eigenvalues}. This happens at the $\Gamma$ point
for the four bands $E^{\alpha}_{-}(\vec k)$ and $E_0(\vec k)$ when they are
folded at $(k_x,k_y)=(2\pi/a,0)$.  If we include the state $|s^{(1)}\rangle$
this becomes a $5 \times 5$ matrix, which cannot be analytically
diagonalized. However, because the state $|s^{(1)}\rangle$ is well below
the Fermi level such that $V_{sp}/(E^{(2)}_{p_{xy}}-E^{(1)}_s) \sim 0.1$, we can apply perturbation theory in $V_{sp}$. More precisely, we will
apply quasi-degenerate stationary perturbation theory\cite{ManousakisQM} to
integrate out the $s$ orbital in second order.
Using Eq.~\ref{eigenstates} for the eigenstates,
the second-order-corrected diagonal matrix-elements (i.e., which include
the effect of the virtual transition from the $|\psi^{(-)}\rangle$ state to $|S^{(1)}\rangle$ state and back) are given as follows:
\begin{eqnarray}
 {\cal E}^{(x)}_{-}(\vec k) &=& E^{(x)}_{-}(\vec k)
 + {{|V^{(x)}_1 V_{sp} d^{\prime}_x|^2} \over {|\xi^{(x)}_{-}|^2
     (E^{(x)}_{-}-E^{(1)}_s)}},\\
 {\cal E}^{(y)}_{-}(\vec k) &=& E^{(y)}_{-}(\vec k)
 + {{|V^{(y)}_1 V_{sp} d^{\prime}_y|^2} \over {|\xi^{(y)}_{-}|^2
     (E^{(y)}_{-}-E^{(1)}_s)}},\\
 {\cal E}^{(x)}_{0}(\vec k) &=& E^{(x)}_{0}(\vec k)
 + {{|V^{(x)}_1 V_{sp} d^{\prime}_x|^2} \over {|\eta^{(x)}|^2
     (E^{(x)}_{0}-E^{(1)}_s)}},\\
 {\cal E}^{(y)}_{0}(\vec k) &=& E^{(y)}_{0}(\vec k)
 + {{|V^{(y)}_1 V_{sp} d^{\prime}_y|^2} \over {|\eta^{(y)}|^2
     (E^{(y)}_{0}-E^{(1)}_s)}}.
\end{eqnarray}
The off-diagonal matrix elements vanish along the high-symmetry directions $\Gamma\to Z \to T \to U \to \Gamma \to X \to U \to \Gamma$ in our plot of
Fig.~\ref{model-with-sp}. Along other directions, the off-diagonal
elements do not necessarily vanish. In this case we would need
to diagonalize the $4 \times 4$ matrix. However, this cannot be done analytically, and if we need to resort to a numerical diagonalization we might as well
diagonalize the full matrix to obtain the more exact non-perturbative solution.
The purpose of this subsection was to demonstrate that the origin
of the dispersion of the upper valence band is from the $V_{sp}$ coupling
which, we feel, has been achieved.

\subsection{Fitting the full crystal with the same tight-binding model}
\label{full-crystal-fit}

Now that we have an analytical understanding of the origin of the band character near the Fermi level,
we can proceed and modify the parameters of our TB model (Table~\ref{table:1}) listed in Table~\ref{table:2} in order
to provide a good fit of the material of our focus, i.e., (BA)$_2$PbI$_4$.

\begin{figure}[!htb]
\includegraphics[width=1.0\linewidth]{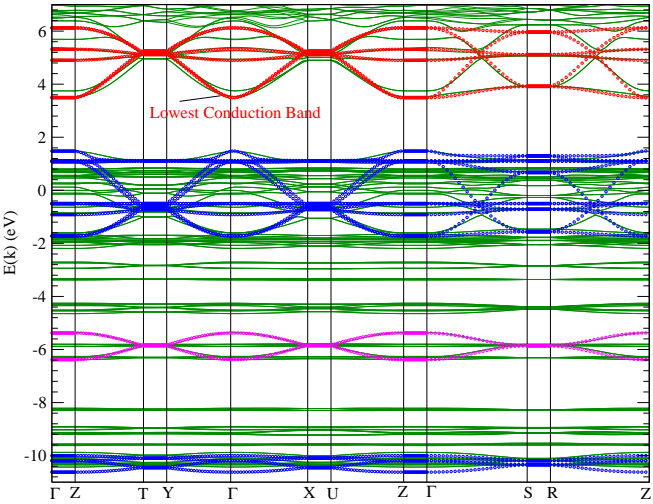} 
\caption{Comparison of the bands of the real material (BA)$_2$PbI$_4$ obtained
  with the DFT (green solid-lines) with the result of the fit using our
  TB model (open circles).}
\label{nnfit}
\end{figure}

The result of the fit is illustrated in Fig.~\ref{nnfit} and the values of the parameters that produce this
TB fit are given in Table~\ref{table:3}.

\begin{table}[!htb]

  \begin{tabular}{ |c|c|c|c|c|c|c|c|c|c|c|}
    \hline
    $E^{(1)}_s$ & $E^{(1)}_{p_{xy}}$ & $E^{(1)}_{p_{z}}$ &
    $E^{(2)}_s$ & $E^{(2)}_{p_{xy}}$ & $E^{(2)}_{p_{z}}$  & $V_{pp\sigma}$ & $V_{pp\pi}$ &
    $V_{sp}$ & $V_{ss}$ & $V_{ps}$ \\
    \hline
    -6.0 & -3.3 & -4.9 & -10.0 & 1.1 & -0.5 &  -1.8 & 0.5 & 1.8 & 0.6 & 0.8 \\
\hline    
  \end{tabular}
  \caption{TB parameters after fitting the DFT bands of the (BA)$_2$PbI$_4$.
    The fit is shown in Fig.~\ref{nnfit}. }
		\label{table:3}

\end{table}
We notice that the TB model describes very well all the bands which are due to the orbitals in PbI$_4$.
However, these orbitals are the only ones that have non-negligible contribution to the bands in the energy range of
out interest, which is discussed in the abstract.

\section{Spin-orbit coupling}
Once we have established our TB model,
it is straightforward to include the contribution of spin-orbit-coupling (SOC).
 The $6s$ orbitals of Pb and the $5s$ orbital
of the two I atoms in the unit cell give no contribution.
The non-zero contribution comes from the $6p$  Pb orbital  and the $5p$ orbital
of the two I atoms (atoms 2 and 3 in our unit-cell). We need to add the following part to the TB Hamiltonian and carry out the diagonalization in
a space of double dimension.
\begin{eqnarray}
  \hat H_{SOC} = J_{\mathrm{Pb}} {\vec \sigma_1} \cdot {\vec l_1} 
+ J_{\mathrm{I}} {\vec \sigma_2} \cdot {\vec l_2} +
J_{\mathrm{I}} {\vec \sigma_3} \cdot {\vec l_3}
\end{eqnarray}
where $J_{\mathrm{Pb}}$ (and $J_{\mathrm{I}}$) is the SOC of the Pb $6p$ (and of the I $5p$)  orbital with the electron spin.  Each of the three terms above in the basis
$| p^{(i)}_x \uparrow\rangle $, $| p^{(i)}_y \uparrow \rangle$, $| p^{(i)}_z \uparrow \rangle$, $| p^{(i)}_x \downarrow \rangle$, $| p^{(i)}_y \downarrow \rangle$, $| p^{(i)}_z \downarrow \rangle$, for $i=1,2,3$ is
given  in Table~\ref{table:4} in units of $J_{i}$ ($J_1 = J_{\mathrm{Pb}}$
, $J_2 = J_{\mathrm{I}}$, $J_3 = J_{\mathrm{I}}$).
\begin{table}[!htb]
  \begin{tabular}{ |c|c|c|c|c|c|c|}
    \hline
& $| p^{(i)}_x \uparrow \rangle$ & $| p^{(i)}_y \uparrow \rangle$ & $| p^{(i)}_z \uparrow \rangle$ & $| p^{(i)}_x \downarrow \rangle$ & $| p^{(i)}_y \downarrow \rangle$ & $| p^{(i)}_z \downarrow \rangle$ \\
    \hline
    $\langle p^{(i)}_x \uparrow |$ & 0 & -$i$ & 0 & 0 & 0 & 1 \\
    \hline
    $\langle p^{(i)}_y \uparrow |$ & $i$ & 0 & 0 & 0 & 0 & -$i$ \\
    \hline
    $\langle p^{(i)}_z \uparrow |$ & 0 & 0 & 0 & -1& $i$ & 0 \\
    \hline
    $\langle p^{(i)}_x \downarrow |$ & 0 & 0 & -1 & 0 & $i$ & 0 \\
    \hline
    $\langle p^{(i)}_y \downarrow |$ & 0 & 0 & -$i$ & -$i$& 0 & 0 \\
    \hline
    $\langle p^{(i)}_z \downarrow |$ & 1 & $i$ & 0 & 0 & 0 & 0  \\
\hline    
  \end{tabular}
  \caption{The SOC Hamiltonian in units of $J_{i}$. }
		\label{table:4}

\end{table}

\begin{figure}[!htb]
\includegraphics[width=1.0\linewidth]{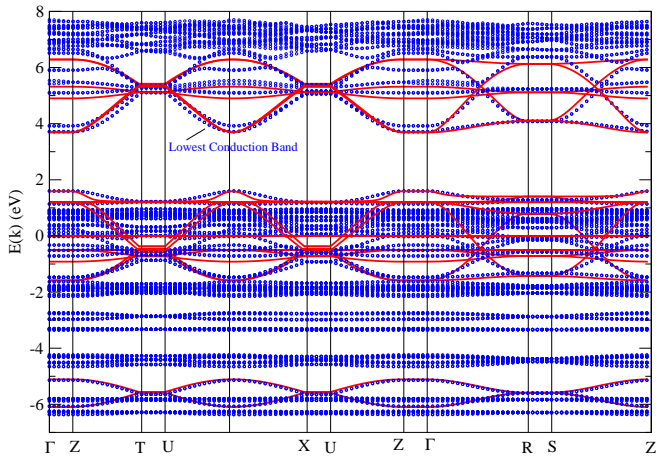} 
\caption{Calculation with fully relativistic pseudopotentials provided in Ref.~\cite{PhysRevB.88.085117} but without
  SOC.}
  \label{fit-no-soc}
\end{figure}

\begin{figure}[!htb]
\includegraphics[width=1.0\linewidth]{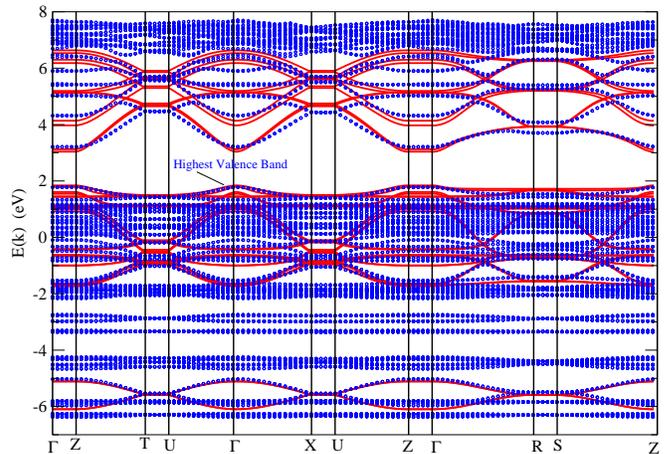} 
\caption{Calculation with  fully relativistic pseudopotentials provided in Ref.~\cite{PhysRevB.88.085117} and with SOC.
  The red lines are the fit using the TB model without changing the values
  of the parameter used in Fig.~\ref{fit-no-soc}, i.e., the parameters used
without SOC, and $\hbar J_1 =0.9 eV$ and $\hbar J_2 =0.7 eV$.}
  \label{soc}
\end{figure}

In order to add the SOC in the DFT calculation we need fully relativistic pseudopotentials (FRP) and we will utilize those provided in Ref.~\cite{PhysRevB.88.085117}.
First, we carry out a SCF calculation on the same size k-point mesh as in our previous calculations and the highest
energy-cutoff suggested in the pseudopotential files.
In Fig.~\ref{fit-no-soc} we illustrate that our TB model fits with the same
level of accuracy the results using these FRP without SOC.
We need this step in order to carry out the fit of the calculation with
SOC. We note that the values of the TB parameters are only slightly off
using this FRP. Next, without changing any of these TB parameters, we simply
add the SOC Hamiltonian described in the previous paragraph and we fit
the results by using only the two fitting parameters $J_{\mathrm{Pb}}$ and
$J_{\mathrm{I}}$. The results are shown in Fig.~\ref{soc}.
The effect of the SOC coupling is large as expected for Pb and I, however,
we also know from previous work that DFT tends to overestimate
the effects of the SOC (see related discussion in Refs.~\cite{PhysRevB.96.165134,PhysRevB.99.035123}). However, the goal of the present paper
is to provide a reasonable starting one-electron model Hamiltonian without
the inclusion of SOC and other effects, such as the effects of
correlations. As discussed, the model can be the starting point
for calculations to include these effects more accurately.
For example, the effect of the SOC could be more accurately included
by fitting the value of the SOC (using the above model independent form)
 to the results of a quasiparticle-self-consistent GW
 calculation\cite{PhysRevB.91.115105} or to the experimental results for the gap or other experimentally determined parameters.

The simple calculation presented above demonstrates the value of the present paper where the TB and the effective
models were determined. Namely, without changing the parameters of the
TB Hamiltonian, i.e., using their values determined without the inclusion of
the SOC, we were able to accurately include the effects of the SOC by simply
adding the SOC to our model.
Similarly, other effects, such as the effects of correlations, the Jahn-Teller effect, optical response, etc,
can be included  starting from
the present model.

\section{Other terms}

There are other important physical effects which yield corrections to the above treatment, such as
the octahedra distortions. 

The octahedral distortions  break the
C$_4$ symmetry and that folds the BZ back to its observed form.
In addition, they open 
  gaps at high-symmetry points  and
  lift band-degeneracies along high-symmetry directions.

  Depending on the problem at hand to address, these terms can be important
  to include, which can be added on top the TB Hamiltonian considered in the
  present work. There are problems, however, where the TB-treatment of the
  present paper can be a good starting point.   For example,
  and this is one of our future projects, starting from this TB-model
  we can include the role of electron interactions to study exciton
  bound-states. These involve bound-states of electron/hole pairs
  excited from near the VBM to near the CBM  where the TB description is
  reasonably good.

  The goal of the present paper was to provide an as simple as possible
  and yet accurate analytic and semianalytic description of the complex
  band-structure of the
  simplest member of the series of the 2D halide-perovskite materials.
  Future work should extend this treatment to $m>1$ members of this
  family and should include the role of the above smaller effects. Another
  direction should be to include the role
of electron interactions in many-body phenomena, such
as the role of excitons in these systems.

\section{Discussion and summary}
\label{discussion}

Simplifying the very complex band-structure of the 2D Ruddlesden-Popper perovskite materials and providing
a simple model which accurately reproduces its main features and, which quantitatively describes it, 
is the main goal of the present paper. Such a simplified picture, not only allows us to
grasp the physics of the electronic structure of these materials, which might help our
thinking forward, but it can also provide a simple description of the one-body part
of an effective many-body Hamiltonian to use to carry out many-body calculations.

First, we have illustrated that, in the simplest case of the series
(BA)$_2$(MA)$_{\mathrm{m}-1}$Pb$_{\mathrm{m}}$I$_{3\mathrm{m}+1}$ with m=1, i.e.,
for (BA)$_2$PbI$_4$,
the bands  in the energy range:  1 eV below the VBM
to 2 eV above the CBM, a 5 eV range covering the range of the photo-electric response,
have negligible contribution from the atomic orbitals contained
in the BA ligands. This conclusion is not to diminish the significance
of organic chains in general.
As an example of their significance, we would like to mention that
the small organic chains, such as the MA, (which are absent in the m=1 case)
reduce dielectric screening  within a monolayer of
RP perovskite materials, which helps generate stable excitons at
room temperature with binding energies of the order of
hundreds of meV\cite{Blancon2020,Quan2019,doi:10.1063/5.0059441}. In addition,
their physical properties  are influenced by the number of layers that affects
the exciton binding energy.
However, in the m=1 case, the role of the atoms of these BA chains
is solely to stabilize the structure
and to act as a charge reservoir which fills completely the highest
occupied bands making the material an insulator. We demonstrate that
other materials, which share the same halide-perovskite core matrix
or even just the ``matrix'' formed by the same halide-perovskite layer,
(keeping the structure and all the atomic distances the same)
have very similar band-structure in the above defined
energy range.

Further, we  analyzed the complex band-structure of this class  of
materials for m=1 and found a simple 2D TB model which can accurately
reproduce the band-structure as
obtained by DFT in the above energy window. We were also able to
simplify this TB model in such a way that it allows an
analytical, transparent and accurate way to describe the origin of all the
features of the band-structure.
As a consequence,  a simple band-picture emerges out of the complexity of
the bands as obtained by straightforward application of DFT.
By considering a two-dimensional TB model, which ignores
the small octahedra distortions, it allows us to reduce the size of
the unit-cell by a factor of two, a fact that doubles the Brillouin-zone
by unfolding it because of symmetry. This reduces the TB Hamiltonian
to a smaller matrix. The most important
conduction and relevant conduction and valence bands are obtained as  a hybridization
of mostly Pb $6p$ and I $5p$ orbitals in the case of our example
(BA)$_2$PbI$_4$. To obtain the correct dispersion
of the highest valence-band, we also needed to involve the role of
the $sp$ hybridization
between the Pb $6s$ orbital and the $5p$ orbital of the I atoms that
form the octahedra corners. Finally, we offer a simple analytic description
of the band-structure by integrating out this Pb $6s$ orbital as it sits
energetically well below the above mentioned energy range.

As already discussed in the previous section, there are several directions
where the results of the present work can be useful and also ways in which
other effects can be incorporated depending on the problem at hand.
For example,  the simple TB model uncovered in this paper will be useful in carrying out many-body calculations to describe
excitonic properties of these materials\cite{PhysRevB.107.075105} which is our future goal.

\section{Acknowledgment}
I would like to thank Hanwei Gao for useful interactions.
This work was supported 
by the U.S. National Science
Foundation under Grant No. NSF-EPM-2110814.

    \appendix
    \section{DFT convergence study}
\label{convergence}
\begin{figure}[!htb]
\includegraphics[width=1.0\linewidth]{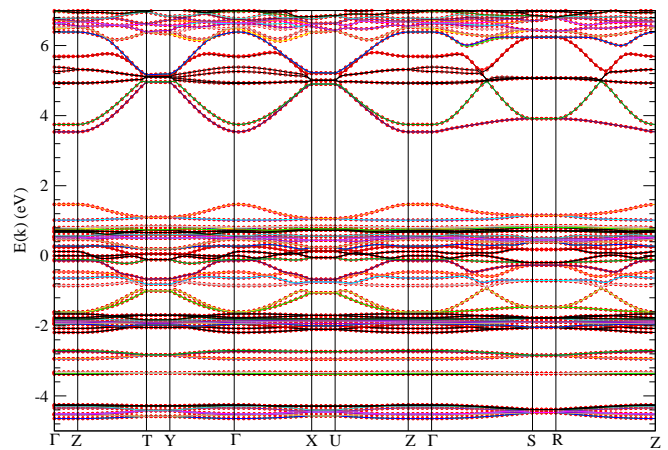} 
\caption{Demonstration of convergence with k-point mesh size.
  The red-open circles are the results obtained on a $8\times 8\times 3$
k-point mesh while the solid lines are the results using the QE default  $4\times 3\times 1$ size.}
\label{comparison-size}
\end{figure}

\begin{figure}[!htb]
\includegraphics[width=1.0\linewidth]{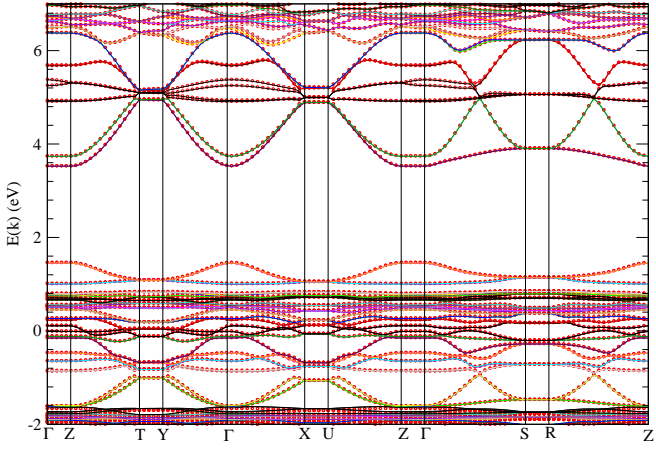} 
\caption{Demonstration of convergence with respect to the energy cut-off $E_c$.
  The red-open circles are the results obtained on a $8\times 8\times 3$
  k-point mesh using $E_c=30 Rd$ while the solid lines are the results using
  $E_c=40 Rd$.}
\label{comparison-cutoff}
\end{figure}

Our self-consistently converged ground-state calculation used a $4\times 3\times 1$  and  a $8\times 4\times 3$
k-point-mesh size. The results are compared in Fig.~\ref{comparison-size}  and we conclude that
the $8\times 6\times 3$ size is large enough for the purpose of the present paper.  In Fig.~\ref{comparison-cutoff}  we compare the results of our DFT
calculation for $8\times 6\times 3$ size for energy cutoff of 30 Rd (red circles)  and 40 Rd (solid lines) to show that using 30 Rd as the energy cutoff is
large enough for the purpose of the present paper.

\section{More complex tight-binding matrix}
\label{tight-binding20}
    Generalization of our TB model of Table~\ref{table:1} for
    the model complete 2d-halide-perovskite illustrated in the right panel of
    Fig.~\ref{reduced-structure} leads to the $20 \times 20$ TB
    matrix given in Table~\ref{table:5}. This model includes the pair of
    I atoms per Pb atom which are off the plane and complete the octahedra.
    \begin{table*}
\scriptsize
\
		\begin{tabular}{ |c|c|c|c|c|c|c|c|c|c|c|c|c|c|c|c|c|c|c|c|c|}
			\hline
  & $\scriptscriptstyle 
s^{(1)}$ & $\scriptscriptstyle p_x^{(1)}$ & $\scriptscriptstyle p_y^{(1)}$ & $\scriptscriptstyle p_z^{(1)}$ & $\scriptscriptstyle s^{(2)}$ & $\scriptscriptstyle p_x^{(2)}$ & $\scriptscriptstyle p_y^{(2)}$ & $\scriptscriptstyle p_z^{(2)}$ & $\scriptscriptstyle s^{(3)}$ & $\scriptscriptstyle p_x^{(3)}$ & $\scriptscriptstyle p_y^{(3)}$ & $\scriptscriptstyle p_z^{(3)}$ & $\scriptscriptstyle s^{(4)}$ & $\scriptscriptstyle p_x^{(4)}$ & $\scriptscriptstyle p_y^{(4)}$ & $\scriptscriptstyle p_z^{(4)}$  & $\scriptscriptstyle s^{(5)}$ & $\scriptscriptstyle p_x^{(5)}$ & $\scriptscriptstyle p_y^{(5)}$ & $\scriptscriptstyle p_z^{(5)}$\\

			\hline

$\scriptscriptstyle  s^{(1)} $ & $\scriptscriptstyle E^{(1)}_{s}$ & 0 & 0 & 0 & $\scriptscriptstyle V_{ss} d_x$ & $\scriptscriptstyle V_{sp} d^{\prime}_x$ & 0 & 0 & $\scriptscriptstyle V_{ss} d_y$ & 0  & $\scriptscriptstyle V_{sp} d^{\prime}_y$ & 0 & $\scriptscriptstyle V_{ss}$ & 0& 0& $\scriptscriptstyle V_{sp}$ & $\scriptscriptstyle V_{ss}$&0 &0 & $V_{sp}$\\

			\hline
$\scriptscriptstyle  p_x^{(1)} $ & 0 & $\scriptscriptstyle E^{(1)}_{p_{xy}}$  & 0 & 0 & $\scriptscriptstyle -V_{ps} d^{\prime}_x$ &$\scriptscriptstyle V_{pp\sigma}d_x$ & 0 & 0 &0&$\scriptscriptstyle V_{pp\pi}d_y$  &0 &0 &0 & $\scriptscriptstyle V_{pp\pi}$& 0& 0&0 & $\scriptscriptstyle V_{pp\pi}$& 0& 0\\

			\hline
$\scriptscriptstyle  p_y^{(1)} $ & 0 & 0 & $\scriptscriptstyle E^{(1)}_{p_{xy}}$& 0 & 0 &0& $\scriptscriptstyle V_{pp\pi}d_x$ &0 & $\scriptscriptstyle -V_{ps}d^{\prime}_y$ &0 & $\scriptscriptstyle V_{pp\sigma} d_y$& 0 &0 & 0& $\scriptscriptstyle V_{pp\pi}$ & 0&0 & 0& $\scriptscriptstyle V_{pp\pi}$ & 0\\

			\hline
$\scriptscriptstyle  p_z^{(1)} $ & 0 & 0 & 0 & $\scriptscriptstyle E^{(1)}_{p_z}$&0 &0 & 0&$\scriptscriptstyle V_{pp\pi} d_x$ & 0 &0 & 0  & $\scriptscriptstyle V_{pp\pi} d_y$ & -$\scriptscriptstyle V_{ps}$ &0 & 0& $\scriptscriptstyle V_{pp\sigma} $& -$\scriptscriptstyle V_{ps}$ &0 & 0& $\scriptscriptstyle V_{pp\sigma} $ \\

			\hline
$\scriptscriptstyle  s^{(2)}$& $\scriptscriptstyle V_{ss}d_x$ & $\scriptscriptstyle V_{ps}d^{\prime}_x$ & 0 &0 & $\scriptscriptstyle E^{(2)}_{s}$ & 0& 0 &0 &0 &0 & 0& 0  & 0&0 &0 &0  &0 &0 &0 &0\\

			\hline
$\scriptscriptstyle  p_x^{(2)} $ &-$\scriptscriptstyle V_{sp}d^{\prime}_x$ & $\scriptscriptstyle V_{pp\sigma} d_x$&0 & 0&0 & $\scriptscriptstyle E^{(2)}_{p_{xy}}$ &0 &0 &0 &0 & 0  &0  & 0 &0 &0 &0&0 &0 &0 &0\\

			\hline
$\scriptscriptstyle  p_y^{(2)} $ &0 &0 &$\scriptscriptstyle V_{pp\pi}d_x$ & 0 &0 &0 & $\scriptscriptstyle E^{(2)}_{p_{xy}}$&0 &0 & 0 & 0 &0  &0 & 0& 0& 0&0 & 0& 0&0\\

			\hline
                        $\scriptscriptstyle  p_z^{(2)} $ &0 &0 &0 &$\scriptscriptstyle V_{pp\pi}d_x$ & 0 &0 &0 &
                        $\scriptscriptstyle E^{(2)}_{p_{z}}$&0 &0 & 0 & 0  &0 & 0& 0&0&0 &0 &0 &0\\
                        
			\hline
$\scriptscriptstyle  s^{(3)} $ &$\scriptscriptstyle V_{ss}d_y$ &0 &$\scriptscriptstyle V_{ps}d^{\prime}_y$ &0 &0 &0 &0 &0 &$\scriptscriptstyle E^{(2)}_{s}$ &0 &0 &0 &0 &0 & 0&0 &0 &0 & 0&0\\

			\hline
$\scriptscriptstyle  p_x^{(3)} $ & 0 &$\scriptscriptstyle V_{pp\pi} d_y$ &0 &0 &0 &0 &0 &0 &0 &$\scriptscriptstyle E^{(2)}_{p_{xy}}$ &0 &0 &0 &0 & 0& 0&0 &0 &0 &0\\

			\hline
$\scriptscriptstyle  p_y^{(3)} $ &$\scriptscriptstyle -V_{sp}d^{\prime}_y$ & 0& $\scriptscriptstyle V_{pp\sigma}d_y$&0 &0 &0 &0 &0 &0 &0 &$\scriptscriptstyle E^{(2)}_{p_{xy}}$ & 0 & 0&0 &0 &0& 0&0 &0 &0\\

			\hline
$\scriptscriptstyle  p_z^{(3)} $ &0 &0 &0 &$\scriptscriptstyle V_{pp\pi}d_y$ &0 &0 &0 &0 &0 &0 &0 & $\scriptscriptstyle E^{(2)}_{p_z}$ &0 & 0& 0&0& 0&0 &0 &0\\

			\hline
$\scriptscriptstyle  s^{(4)} $ &$\scriptscriptstyle V_{ss}$ & 0 & 0&$\scriptscriptstyle V_{ps}$  &0 &0 &0 &0 &0 &0&0 &0 &$\scriptscriptstyle E^{(2)}_{s}$ &0 &0 &0 & 0&0 &0 &0\\

			\hline
$\scriptscriptstyle  p_x^{(4)} $ & 0 &$\scriptscriptstyle V_{pp\pi} $ &0 &0 &0 &0 &0 &0 &0 & 0 & 0& 0&0 &$\scriptscriptstyle E^{(2)}_{p_{xy}}$ &0 &0  & 0& 0&0 &0\\

			\hline
$\scriptscriptstyle  p_y^{(4)} $ & 0 & 0& $\scriptscriptstyle V_{pp\sigma}$&0 &0 &0 &0& 0 &0 &0 &0 & 0&0 &0 &$\scriptscriptstyle E^{(2)}_{p_{xy}}$ & 0 &0 &0 &0 &0\\

			\hline
$\scriptscriptstyle  p_z^{(4)} $ &-$\scriptscriptstyle V_{sp} $ &0 &0 &$\scriptscriptstyle V_{pp\pi}$ &0 &0 &0 &0 &0 &0 &0 & 0&0 &0 &0 &$\scriptscriptstyle E^{(2)}_{p_z}$  &0 &0 &0 &0\\

			\hline
$\scriptscriptstyle  s^{(5)} $ &$\scriptscriptstyle V_{ss}$ & 0 & 0&$\scriptscriptstyle V_{ps}$  &0 &0 &0 &0 &0 &0 &0&0&0&0&0&0 &$\scriptscriptstyle E^{(2)}_{s}$ &0 &0 &0\\

			\hline
$\scriptscriptstyle  p_x^{(5)} $ & 0 &$\scriptscriptstyle V_{pp\pi} $ &0 &0 &0 &0 &0 &0 &0 & 0 & 0& 0&0 &0&0&0&0&$\scriptscriptstyle E^{(2)}_{p_{xy}}$ &0 &0 \\

			\hline
$\scriptscriptstyle  p_y^{(5)} $ & 0 & 0& $\scriptscriptstyle V_{pp\sigma}$&0 &0 &0 &0& 0 &0 &0 &0 & 0&0 &0 &0&0&0&0&$\scriptscriptstyle E^{(2)}_{p_{xy}}$ & 0 \\

			\hline
$\scriptscriptstyle  p_z^{(5)} $ &-$\scriptscriptstyle V_{sp} $ &0 &0 &$\scriptscriptstyle V_{pp\pi}$ &0 &0 &0 &0&0&0&0&0 &0 &0 &0 & 0&0 &0 &0 &$\scriptscriptstyle E^{(2)}_{p_z}$ \\

			\hline
		\end{tabular}
		\caption{TB matrix when the pair off-the-plane I atoms are included. }
		\label{table:5}

\end{table*}

\end{document}